\documentclass[ ]{aastex631}
\usepackage{amssymb}
\usepackage{amsmath}
\usepackage{appendix}

\newcommand{\K}{\text{ K}}
\newcommand{\eV}{\text{ eV}}

\newcommand{\cm}{\text{ cm}}
\newcommand{\g}{\text{ g}}
\newcommand{\s}{\text{ s}}

\makeatletter

\newcommand{\Rmnum}[1]{\expandafter\@slowromancap\romannumeral #1@}
\makeatother

\shortauthors{Chen $\&$ Dai}
\shorttitle{EM Counterparts of BBH mergers in AGN disk}

\begin{document}

\title{Electromagnetic Counterparts Powered by Kicked Remnants of Black Hole Binary Mergers in AGN Disks}

\author[0000-0001-8955-0452]{Ken Chen}

\affiliation{School of Astronomy and Space Science, Nanjing University, 
	Nanjing 210023, China}

\author[0000-0002-7835-8585]{Zi-Gao Dai}

\affiliation{Department of Astronomy, School of Physical Sciences, 
	University of Science and Technology of China, Hefei 230026, China; daizg@ustc.edu.cn}

\begin{abstract}
The disk of an active galactic nucleus (AGN) is widely 
regarded as a prominent formation channel 
of binary black hole (BBH) mergers that can 
be detected through gravitational waves (GWs). 
Besides, the presence of dense environmental gas 
offers the potential for an embedded BBH merger 
to produce electromagnetic (EM) counterparts. 
In this paper, we investigate EM emission 
powered by the kicked remnant of a BBH merger 
occurring within the AGN disk. The remnant BH will 
launch a jet via accreting magnetized medium 
as it traverses the disk. The resulting jet 
will decelerate and dissipate energy into a 
lateral cocoon during its propagation. 
We explore three radiation mechanisms
of the jet-cocoon system: jet breakout emission, 
disk cocoon cooling emission, and jet cocoon cooling 
emission, and find that the jet cocoon cooling 
emission is more likely to be detected in its own 
frequency bands. We predict a soft X-ray transient, 
lasting for $O(10^3)\s$, to serve as an EM counterpart,
of which the time delay $O(10)~days$ after the GW trigger 
contributes to follow-up observations. Consequently,
BBH mergers in the AGN disk represent a novel multimessenger
source. In the future, enhanced precision in 
measuring and localizing GWs, coupled with diligent 
searches for such associated EM signal, will effectively 
validate or restrict the origin of BBH mergers in the AGN disk. 
\end{abstract}


\section{Introduction}
 
Since the first observation of gravitational 
wave (GW) by the advanced 
LIGO/Virgo GW detectors \citep{Abbott16}, 
binary black hole (BBH) mergers are emerging 
as a predominant category of GW sources 
\citep{Abbott19, LIGO21a, LIGO21b}. Future GW detectors 
are expected to detect a substantial number (e.g., 
$\sim 10^4-10^5$ events per year) of BBH mergers 
\citep{Baibhav19, Saini23}. 
Various formation channels have been proposed to 
account for BBH mergers 
(\citealt{Mapelli21, Mandel22}; 
see \citealt{Tagawa20b} for a list). 
Among these channels, the disk of an 
active galactic nucleus (AGN) 
is garnering significant attention in 
recent studies, as the BBH merger event GW190521 
\citep{Abbott20a}, along with a plausible 
electromagnetic (EM) counterpart ZTF19abanrhr 
\citep{Graham20}, indicates an AGN disk 
environment origin. 

Black hole binaries are expected to widely exist
in the AGN disk \citep[e.g.][]{Bartos17, Stone17, McKernan20a,
McKernan20b, Tagawa20a, LiJ22, LiJ23, DeLaurentiis23}. 
Several mechanisms, e.g., 
gas torque \citep[e.g.][]{LiY21, LiY22, LiR22, LiR23, Rowan23}, 
binary-single interaction 
\citep[e.g.][]{Leigh18, Tagawa20a, Samsing22}, and 
GW radiation \citep{Shapiro83}, can induce
orbital hardening, ultimately leading to BBH 
mergers. Due to the unique nature of an AGN (strong 
gravity from a central supermassive black hole, 
high stellar density in the nuclear cluster, and 
the presence of an accretion disk), BBH mergers in 
the AGN disk exhibits several peculiarities, such 
as hierarchical mergers \citep[e.g.][]{Yang19a,Tagawa21b} or 
eccentric mergers \citep[e.g.][]{Tagawa21a, Samsing22}. The 
merger rate of BBH with an AGN disk origin 
ranges from $O(10^{-3})$ to 
$O(10^2)~\rm{Gpc}^{-3}~\rm{yr}^{-1}$ 
(see Table 1 in \citealt{Arca Sedda23}). Despite 
significant uncertainty, these mergers 
can make a substantial contribution to 
the overall population with the inferred 
redshift-dependent merger rate
$17.9-44~\rm{Gpc}^{-3}~\rm{yr}^{-1}$ 
\citep{Abbott23}. Therefore, the AGN disk formation  
channel demonstrates great importance in the era 
of GW astronomy.

In addition to the distinctive GW signals, the  
gas-rich environment of AGN disk provides an 
opportunity for the embedded BBHs to generate 
electromagnetic (EM) emission, elevating these 
BBHs to multimessenger sources. In turn, the 
identification of an EM counterpart would 
strongly support the AGN disk formation channel
\citep[e.g.][]{Veronesi22, Veronesi23}.
Preliminarily, in terms of energy, EM emission 
from the binaries is potentially observable,
as the total power of energy release during 
the BBH accretion process exceeds the AGN 
background luminosity \citep{Stone17, Bartos17}. Various 
specific emission processes have been proposed
to predict the EM counterparts \citep{McKernan19, 
Graham20, Kimura21, Wang21b, Tagawa23a, Tagawa23b,
Rodriguez-Ramirez23}, with diverse emission 
properties across different models. Though 
several candidates are reported \citep{Graham23}, 
no EM counterpart to BBH mergers in the AGN 
disk has been confirmed so far, thus preventing a 
test of existing models and leaving room 
for new ones.

In this paper, we investigate a plausible 
emission process by concretely considering the 
accretion of BHs both during the inspiral phase 
and post-merger. In our model, a jet launched 
by the kicked merger remnant BH acts as an 
emission source. We study the propagation 
of the jet within AGN disk and the subsequent 
emergence of resulting emission, with
a specific focus on its detectability.\footnote
{Note that after the submission of this 
paper, \cite{Tagawa23b} was posted online, which 
also investigates the cocoon cooling emission resulting 
from a jet produced by the BBH merger remnant embedded
in an AGN disk. The differences lie in the treatment of 
mass rate accreted onto the merger remnant and the
time delay between GW trigger and EM emission. In 
\cite{Tagawa23b}, a jet launch occurs immediately after 
the merger; in our study, a jet would be launched 
following a sufficient accumulation of magnetic 
field around the kicked remnant BH during its traversal 
through AGN environment. By combining emissions with various 
time delays, a more distinctive EM counterpart can be 
identified.}
This paper is organized as follows. In Section 
\ref{section2}, we discuss the circum-binary
environment of an embedded BBH merger and the 
properties of GW recoil kick impacted on the 
remnant. We investigate the formation and 
propagation of a jet driven by the kicked 
remnant in Section \ref{section3}. We next 
explore three emission processes of the 
jet-cocoon system in Section \ref{section4}. 
Several discussions are presented in Section
\ref{section5}, particularly involving the 
detectability of associated EM counterparts 
in Section \ref{Detectability}. We summarize 
our conclusions in Section \ref{section6}. 
Symbols $c$ and $G$ in this paper denotes 
the speed of light and the gravitational 
constant, respectively.

\section{BBH merges and kicked in AGN disk environment}
\label{section2}
The AGN disk can be described as a viscous accretion 
disk with three parameters, i.e., the SMBH mass $M$, 
the mass inflow rate $\dot{M}$, and the viscosity 
parameter $\alpha$ \citep{Shakura73}; meanwhile, the outer region 
of the AGN disk is gravitationally unstable to impact the disk 
structure \citep{Goodman03}. Many plausible theoretical models have 
been proposed to construct a quasi-stationary 
self-gravitating accretion disk \citep[e.g.][et al.]
{Sirko03,Thompson05, Mishra20, Gilbaum22}, among which 
we adopt the widely used SG \citep{Sirko03} model and 
TQM \citep{Thompson05} model in this work, and we refer to 
Equation (26)-(29) of \citep{Pan21a} for disk equations.

\subsection{Circum-binary environment of a BBH merger}
As the gas in an AGN disk is differentially rotating and 
its density is high, a circularized disk would form around the 
embedded BH with initial mass inflow rate greatly exceeding 
the Eddington accretion rate,
$\dot{M}_{\rm{Edd}}=L_{\rm{Edd}} c^2 / \eta$, where
we set the radiation efficiency $\eta=0.1$ and
$L_{\rm{Edd}}=4\pi G m m_pc/\sigma_T=1.26\times10^{40}m_2\ 
\rm{erg\,s}^{-1}$ \citep[e.g.][]{Wang21a, Pan21b, ChenD23}.  
Similarly, a circum-binary disk would form 
around the embedded binary black hole. In addition, two 
circum-BH disks severally form around the two components (the 
accreting structure is visible in recent numerical 
simulations, e.g. \citealt{LiY21, Dempsey22, Kaaz23a}).
We employ the Bondi-Holye-Lyttleton (BHL) formulation 
\citep[e.g.][]{Edgar04}, 
$\dot{M}_{\rm{BHL}}=4 \pi G^2 m_{\rm{bin}}^2 
\rho_{\rm{AGN}} /(v_{\rm{rel}}^{2}+c_{\rm{s,AGN}}^{2} )^{3/2}$,
where $m_{\rm{bin}}$ is the total mass of the BBH, 
$\rho_{\rm{AGN}}$ and 
$c_{\rm{s,AGN}}$ are the density and sound speed
of AGN disk, $v_{\rm{rel}}$ is the relative velocity between 
the BBH center of mass and the ambient gas, to estimate the mass 
captured rate of the BBH system and the initial mass inflow  
rate of the circum-binary disk. For the case of the binary 
separation $a_{\rm{bin}}$ being much smaller than the outer 
boundary of the circum-binary disk $r_{\rm{obd}}$, 
the BBH can be approximated as a 
point source and the circum-binary disk would evolve akin to a 
circum-single disk \citep{Kimura21}, 
though the mass rate would be reduced 
and evolve over time in the BBH system \citep[e.g.][]{LiR22}.

\begin{figure*}
	\begin{center}
		\includegraphics[width=0.45\textwidth]{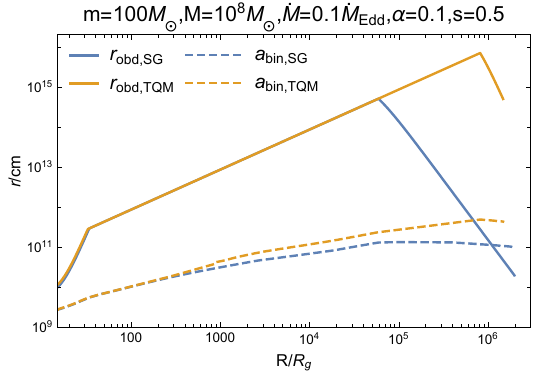}
		\includegraphics[width=0.45\textwidth]{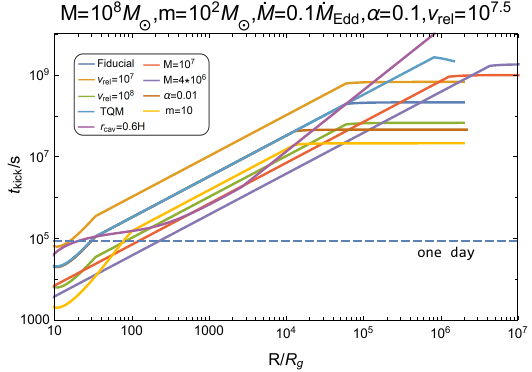}
	\end{center}
	\caption{Properties of binary accretion and merger. The left panel
		shows the outer boundary radius of circum-binary disk $r_{\rm{obd}}$, 
		of which the calculations adopt \cite{ChenD23}, considering both 
		the circularization due to angular momentum conservation of 
		the BBH-captured AGN disk gas
		and the disk truncation by the heated outflow, and 
		the critical binary separation $a_{\rm{bin,cr}}$ as a function of 
		AGN radius $R$, scaled by gravitational radius 
		$R_{\rm{g}}=G M/c^2$; the fiducial parameters are 
		$m_{\rm{bin}}=100\rm{M}_{\odot}$, $M=10^{8}\rm{M}_{\odot}$, 
		$\dot{M}=0.1 \dot{\rm{M}}_{\rm{Edd}}$, $\alpha=0.1$, $s=0.5$, 
		where $s$ is the power-law index of the radius-dependent 
		circum-binary disk mass inflow rate. The blue and golden 
		lines represent the AGN disk adopting SG and TQM model,
		respectively.
		We change all these parameters and find 
		that the relative size between $r_{\rm{obd}}$
		and $a_{\rm{bin,cr}}$ are preserved, 
		i.e., $a_{\rm{bin,cr}} \ll r_{\rm{obd}}$;
		so, for simplicity, only the fiducial case is shown.
	    The right panel represents the timescale of a kicked merger 
	    remnant crossing the cavity, $t_{\rm{kick}}$, 
	    under various system parameters (as shown in
	    the inset); the fiducial case is the same, with 
	    $v_{\rm{rel}} \sim v_{\rm{k}} = 10^{2.5}\rm{~km~s}^{-1}$ 
	    ($\rm{cm~s}^{-1}$ in graph legend) adopting SG model. 
        The blue dashed line represents the time equaling to 
        one day. }
	\label{Fig:abin}
\end{figure*}

As shown by \cite{ChenD23}, a hyper-Eddington circum-BH disk would launch 
powerful outflow to push out the surrounding gas, generating 
a low-density cavity to intermittently choke the 
effective accretion; in the same way, the embedded BBH would go 
through similar accretion processes. Meanwhile, the binary 
separation hardens due to GW radiation within the inspiral 
timescale \citep[e.g.][]{Shapiro83},
\begin{equation}
t_{\rm{GW}}=\frac{5}{128} \frac{c^{5} 
	a_{\rm{bin}}^{4}}{G^{3} m_{\rm{bin}}^{3}}. 
\end{equation}
We compare $t_{\rm{GW}}$ with the cavity 
persistence timescale $\sim t_{\rm{ref}}$, 
Equation (32) in \cite{ChenD23}, 
of each recurrent accretion process. For a BH binary system with 
an initial separation $a_{\rm{bin}}$ satisfying 
$t_{\rm{GW}}(a_{\rm{bin}})>t_{\rm{ref}}$, 
the BBH would not merge within one 
accretion episode, but continuously hardens via various 
mechanisms \citep[e.g.][]{Tagawa20a}; then, when
$a_{\rm{bin}}<a_{\rm{bin,cr}}$, where 
$t_{\rm{GW}}(a_{\rm{bin,cr}})=t_{\rm{ref}}$, the BBH 
would undergo inspiral to merger just within 
one recurrent accretion process.
As shown in Figure \ref{Fig:abin}, the critical binary 
separation $a_{\rm{bin,cr}}$ is much smaller than $r_{\rm{obd}}$, 
thus, the binary can be treated as a single object and the whole 
accretion-feedback process is valid. Also, since
$a_{\rm{bin,cr}}\lesssim O(10^{11})\cm$, the BBH hardening
is dominated by GW radiation, the other mechanisms, 
e.g., gas dynamical friction, binary–single interaction 
and torque from the circum-binary disk, 
can be ignored \citep{Tagawa20a}.

For a BBH merger occurring within one recurrent 
accretion process, the cavity persistence timescale 
is much longer than the timescale of outflow formation to
open a cavity, which 
is estimated by the viscous timescale of the circularized 
circum-binary disk, $t_{\rm{vis}}$, where 
$t_{\rm{vis}}\lesssim0.1t_{\rm{ref}}$ \citep{ChenD23}.
Consequently, it is more probable for the 
merging BBH to reside in a low-density cavity 
rather than being surrounded by AGN disk gas. However, prior to
the GW dominates over BBH hardening, binary-single interactions 
are frequent \citep{Tagawa20a}, leading to an appreciable
probability of the BH binary undergoing a 3-body merger 
\citep{Samsing22}, i.e., the temporary BH binary merges in 
the chaotic 3-body system due to high binary orbital 
eccentricity decreasing the inspiral time. 
No work has been done to investigate the
accretion process of such 3-body system. Although unproven, we 
propose that a stable circum-binary disk and outflow would not 
persist under the perturbation of a chaotic third companion, 
preventing the formation of a large-scale cavity in the 
3-body system, and thereby the merger remnant would be 
encompassed by the AGN disk gas. 
Moreover, the study of accretion-induced
outflow feedback relies on BH (BBH) systems that are co-rotating 
with the AGN disk \citep[e.g.][]{Kimura21, Tagawa22, ChenD23},
if the BH binary has a large velocity relative to the 
ambient gas, it remains unclear whether a stable circum-binary 
disk can form to open a cavity. Therefore, further exploration
is required to elucidate the specific feedback mechanism.

In brief, at the time of a BBH merger 
via GW radiation in the AGN disk, the circum-binary 
environment could be either the AGN disk or the cavity, 
depending on whether a stable disk forms 
around the inspiral BH binary and an accretion-induced 
strong outflow is successfully generated.

\subsection{GW kick on a BBH merger remnant}
For a BBH binary with asymmetric components, the merger remnant 
would receive a recoil kick via anisotropic gravitational
radiation during the final inspiral stage, where the 
kick velocity $v_{\rm{k}}$ depends 
on the binary mass ratio and the BHs' spin 
\citep[e.g.][et al.]{Gonzalez07, 
Campanelli07a, Baker07, Herrmann07}. The velocity
component caused by the mass unequal is \citep{Campanelli07b}
\begin{equation}
v_{\rm{m}}=A \frac{q^{2}(1-q)}{(1+q)^{5}}\left[1+B \frac{q}{(1+q)^{2}}\right], 
\end{equation}
where $q$ is the mass ratio of the lighter to the heavier BH, and
$A=1.2 \times 10^4 \rm{~km~s}^{-1}$, $B=-0.93$. For the 
LIGO/Virgo detected events \citep{Abbott19, LIGO21a, LIGO21b}, 
$v_{\rm{m}}$ of BBH mergers ranges from $43\rm{~km~s}^{-1}$ to 
$175\rm{~km~s}^{-1}$, with an average value $109\rm{~km~s}^{-1}$.
Also, the amplitude of $v_{\rm{k}}$ demonstrates remarkable 
sensitivity to the BH spin and can scale up to 
$O(10^3)\rm{~km~s}^{-1}$ \citep{Campanelli07b}, e.g., for the 
event GW190521 \citep{Abbott20a}, $v_{m}\sim 70 \rm{~km~s}^{-1}$ but 
$v_{\rm{k}} > 200 \rm{~km~s}^{-1}$ when taking the effects 
of BH spin into account \citep{Abbott20b}. 
Since the components of the embedded BBH would inherently  
possess low spin \citep{Chen23}, or spin rapidly, but of 
which the individual spin directions are misaligned with 
the binary orbital angular momentum \citep{Tagawa20b}, 
we set three values of $v_{\rm{k}}= 10^2, 10^{2.5},
10^3\rm{~km~s}^{-1}$ to study the kicked merger remnants of 
BBH with various mass ratio, spin magnitude and direction.

As demonstrated above, a BBH merger would take place 
in a low-density cavity provided that a stable 
circum-binary accretion disk exists during the binary inspiral 
phase; subsequently, the kicked merger remnant need traverse 
the cavity to re-enter the AGN disk environment. The radius
of the cavity, $r_{\rm{cav}}$, is determined by the 
specific properties of the accretion-induced outflow, which 
is about the order of BBH gravity radius 
\citep{Kimura21, ChenD23}, i.e., 
min$\left\lbrace r_{\rm{BHL}},r_{\rm{Hill}}\right\rbrace$, 
where $r_{\rm{BHL}}\simeq G m_{\rm{bin}}/(v_{\rm{rel}}^{2}+c_{\rm{s,AGN}}^{2})$ 
is the BHL radius, and $r_{\rm{Hill}}=(m_{\rm{bin}}/3M)^{1/3}R$
is the Hill radius, or is approximated to $0.6H$ \citep{Tagawa22},
where $H$ denotes the vertical scale height of the disk. We 
estimate the timescale of a kicked remnant crossing the cavity
as
\begin{equation}
	t_{\rm{kick}}=r_{\rm{cav}}/v_{\rm{k}}\simeq 37~days
	\left(\frac{r_{\rm{cav}}}{10^{14}\cm}\right)
	\left(\frac{v_{\rm{k}}}{10^{7.5}\cm\s^{-1}}\right)^{-1}.
\end{equation}
As shown in Figure \ref{Fig:abin},
$t_{\rm{kick}}$ increases with the binary location $R$, which 
would enlarge the time delay between a BH merger and its
associated EM emission, since the BH needs to interact with the 
dense AGN disk gas to efficiently release energy (see below). 
However, given the uncertain existence and size of a cavity 
surrounding the merging BBH, we assume that the remnant 
is directly kicked into the AGN disk environment, and discuss 
the potential time delay caused by cavity in Section \ref{4.4}. 

\section{Interaction between kicked remnant and AGN disk environment --- jet and cocoon }
\label{section3}

\subsection{Formation and power of BH-driven jet}
After the embedded BBH merger, the kicked remnant moving through 
the AGN disk with $v_{\rm{k}}$ would accrete the ambient gas with a 
mass inflow rate $\dot{M}_{\rm{BHL,k}}$, 
\begin{equation}
\dot{M}_{\rm{BHL,k}}=\frac{4 \pi G^2 m_{\rm{bin}}^2 
\rho_{\rm{AGN}}} {(v_{\rm{k}}^{2}+c_{\rm{s,AGN}}^{2} )^{3/2}} 
\simeq1.6 \times 10^7 \dot{M}_{\rm{Edd}}
\left(\frac{m_{\rm{bin}}}{100 M_\odot}\right)
\left(\frac{\rho_{\rm{AGN}}}{10^{-10}\,{\rm g\,cm^{-3}}}\right)
\left(\frac{v_{\rm{k}}}{10^{7}\,{\cm \s^{-1}}}\right)^{-3}.
\end{equation}
where in the second equality, only $v_{\rm{k}}$ is considered 
for an intuitive order of magnitude estimation, while in the subsequent 
specific calculation of $\dot{M}_{\rm{BHL,k}}$, both 
$v_{\rm{k}}$ and $c_{\rm{s,AGN}}$ are taken into account.
The hyper-Eddington accretion is realizable, since for BHL 
accretion, the inflow is highly 
quasi-spherical \citep[e.g.][]{El Mellah15} and 
the released gravitational energy is advected inward with gas 
\citep{Begelman79}. Meanwhile, the AGN disk would be magnetized, 
of which the ratio of ambient gas to magnetic 
pressure $\beta \lesssim 10-100$ 
\citep{King07, Salvesen16, Kaaz23b}. 
Recent numerical simulations indicate that the 
ambient magnetic fields are frozen and flow inward 
with gas, resulting in their accumulation 
to saturation near the BH \citep{Kwan23,Kaaz23b}. Due to a 
large spin of the merger remnant \citep[e.g.][]{Tichy08, McKernan23}, 
a jet would be generated via the Blandford-Znajek (BZ) mechanism 
\citep{Blandford77}, of which the power can be estimated as 
\begin{equation}
	L_{\rm{jet}}=f \dot{M}_{\rm{BHL,k}} c^2,
\end{equation}
where $f \sim 0.1$ for $\beta \lesssim 10-100$ (according to 
Figure 3 in \citealt{Kaaz23b}), containing effects of both 
the jet efficiency and the reduction of accretion rate.

\begin{figure*}
	\begin{center}
		\includegraphics[width=0.5\textwidth]{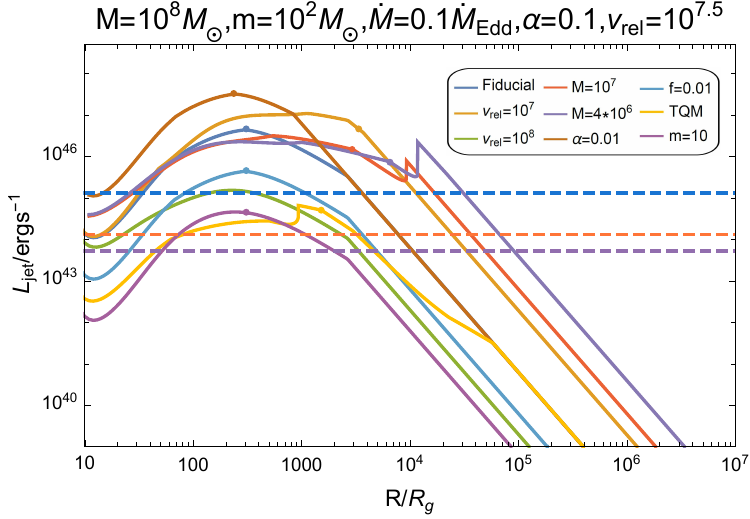}
	\end{center}
	\caption{Power of a jet launched by a kicked merger 
		remnant moving through the magnetized AGN disk 
		environment with different system parameters 
		(the inset shows these parameters). 
		The blue, orange, and purple dashed lines represent the bolometric 
		luminosity of AGN disk $L_{\rm{bol,AGN}}=0.1 \dot{M} c^2$ with 
		$M=10^{8}, 10^{7}, 4\times10^{6}~\rm{M}_{\odot}$, respectively. 
		The point laid on each 
	    curve represents $t_{\rm{BHL}}=t_{\rm{k}}$ with the right side 
        $t_{\rm{BHL}}<t_{\rm{k}}$.}
	\label{Fig:Ljet}
\end{figure*}

As shown in Figure \ref{Fig:Ljet}, overall, the power of 
the BH-driven jet surpasses the bolometric luminosity 
of AGN disks, $L_{\rm{bol,AGN}}$, across an
extensive range of $R$, 
thereby rendering the jet potentially observable 
for efficient specific radiation 
processes. Some cases possess weak jets with 
$L_{\rm{jet}}<L_{\rm{bol,AGN}}$, 
thus unfavorable for observations, e.g., the 
$v_{\rm{k}} = 10^{3}\rm{~km~s}^{-1}$ and 
$m_{\rm{bin}}=10\rm{M}_{\odot}$ case due to a lower 
$\dot{M}_{\rm{BHL,k}}$, and the $\rm{TQM}$ case due to a lower disk 
density compared with the SG model. Also, 
we reduce $f$ to $0.01$ on account of the captured AGN disk
gas likely carrying non-negligible angular momentum, which 
would potentially reduce the mass rate 
accreted onto BH, and find that 
though contracted, there is still a range of $R$ for 
$L_{\rm{jet}}>L_{\rm{bol,AGN}}$.

In addition, since the embedded BBH would undergo binary-single 
interaction to alter its direction of orbital angular momentum 
\citep[e.g.][]{Tagawa20a, Samsing22, Valtonen06}, 
and given that the direction of 
GW kick strongly relies on spin orientations 
\citep[e.g.][]{Campanelli07a},
$\vec{v}_{\rm{k}}$ may be randomized. Consequently, 
the minimum timescale for a 
merger remnant to exit the AGN disk is estimated as 
$t_{\rm{k}}=H/v_{\rm{k}}$.
Meanwhile, the jet formation requires at least a few accretion 
timescales $t_{\rm{BHL}}\simeq r_{\rm{BHL}}/v_{\rm{k}} 
\sim G m_{\rm{bin}}/v_{\rm{k}}^3$ \citep{Kaaz23b}. 
So the merger remnant is 
more feasible to launch jet within AGN disk when 
$t_{\rm{BHL}}<t_{\rm{k}}$, which holds over large 
$R$ regions, as shown in Figure \ref{Fig:Ljet}.

 \begin{figure*}
 	\begin{center}
 		\includegraphics[width=0.5\textwidth]{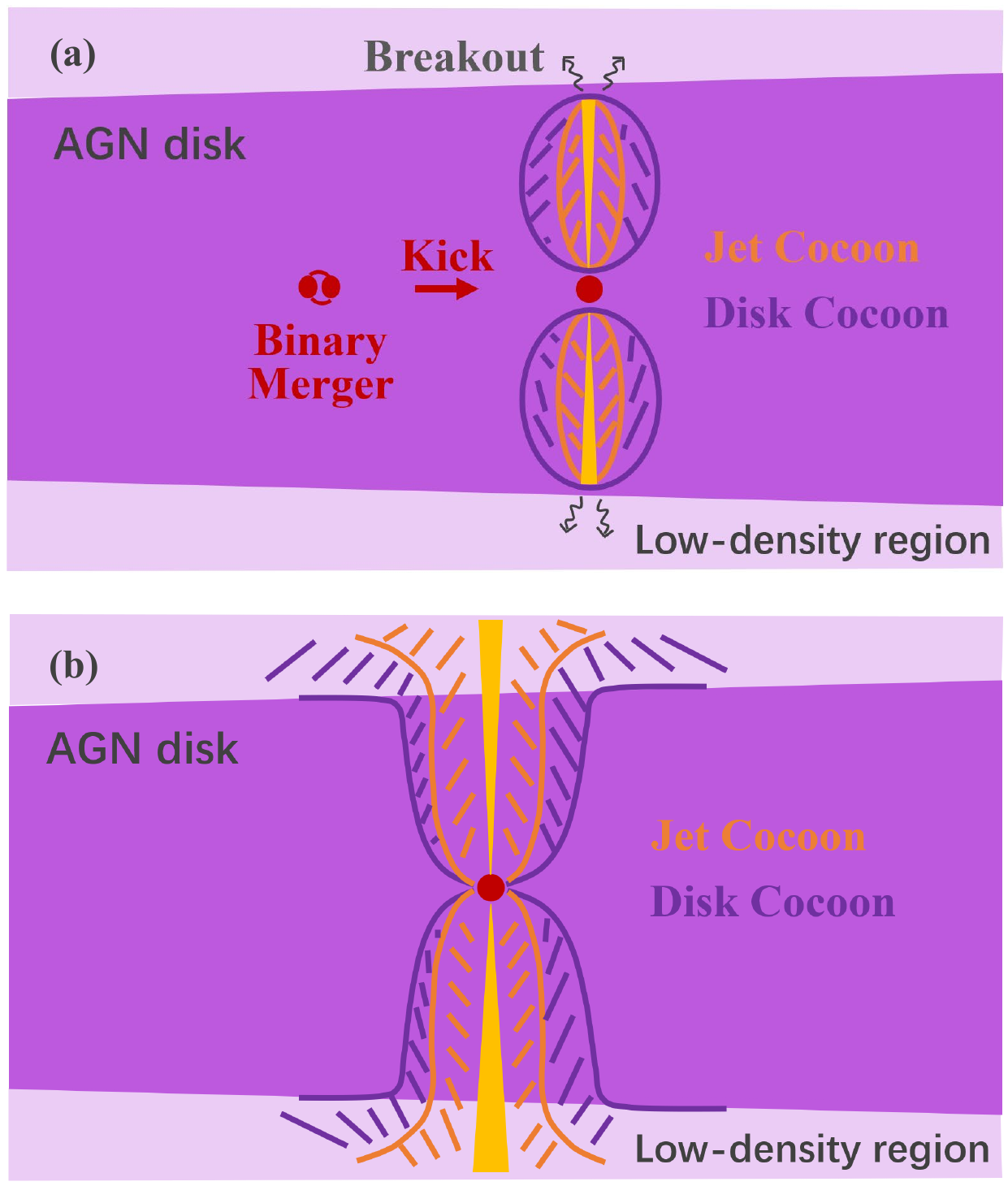}
 	\end{center}
 	\caption{Schematic diagram illustrating 
 		the evolution of a jet and a cocoon
 		driven by a kicked BBH merger 
 		remnant traversing the AGN disk. 
 		(a) The kicked remnant accretes ambient 
 		magnetized gas and launches a jet, which 
 		propagates through and interacts with 
 		the AGN disk to generate a two-component 
 		cocoon; at the jet breakout, photons can 
 		diffuse out from the jet head to produce
 		emission. (b) After the jet breakout, 
 		the jet cocoon and disk cocoon severally 
 		expands, cools, and produces emission.}
 	\label{Fig:sketch}
 \end{figure*}

\subsection{Propagation and breakout of a jet within the AGN disk}
\label{sec-jet-propagation}
We consider a representative scenario where a BBH merger occurs 
at the midplane of the AGN disk, 
and the remnant is kicked along the disk plane. 
Additionally, we assume that a jet driven by the BZ mechanism 
propagates vertically to the disk plane. 
A schematic diagram illustrating the evolution of a 
kicked BH and its associated jet is depicted in 
Figure \ref{Fig:sketch}. The jet interacts with the 
dense ambient medium during its propagation, 
forming a shocked structure at its head position 
and a lateral hot cocoon, which would in 
turn collimate the jet \citep[e.g.][]
{Begelman89, Matzner03, Lazzati05, Bromberg11}. 
The jet is initially Poynting flux dominated, but
would behave as a baryon dominant one at large scale 
due to the early dissipation of magnetic fields 
\citep[e.g.][]{Bromberg16, Nakar17, Gottlieb22}. 
For simplicity, therefore, we consider 
a hydrodynamic jet and investigate its propagation 
through the AGN disk.

We calculate the evolution of the jet and cocoon following the 
descriptions in \cite{Bromberg11}. Collision of the jet with 
the ambient AGN disk gas at its head generates a 
forward shock and a reverse shock, where the ram pressure 
balance sets the velocity of jet head as
\begin{equation}\label{vh}
	\beta_{\rm{h}}=\frac{\beta_{\rm{j}}}{1+\tilde L^{-1/2}},
\end{equation}
where the jet velocity $\beta_{\rm{j}} \sim 1$, and the collimation 
parameter is
\begin{equation}\label{Lcol}
\tilde L\equiv\frac{\rho_{\rm{j}}
	h_{\rm{j}}\Gamma_{\rm{j}}^2}{\rho_{\rm{AGN}}}\simeq\frac{L_{\rm{jet}}}{\Sigma_{\rm{j}}\rho_{\rm{AGN}}c^3},
\end{equation}
where $\rho_{\rm{j}}$, $h_{\rm{j}}$ and $\Gamma_{\rm{j}}$ are density, specific 
enthalpy and Lorentz factor of the jet; the second 
equation holds as the total jet energy density 
$\rho_{\rm{j}}h_{\rm{j}}\Gamma_{\rm{j}}^2-\rho_{\rm{j}}\Gamma_{\rm{j}}-p_{\rm{j}} \sim \rho_{\rm{j}}h_{\rm{j}}\Gamma_{\rm{j}}^2 
=L_{\rm{jet}}/\Sigma_{\rm{j}}c^3$ \citep{Matzner03}, where $p_{\rm{j}}$ and 
$\Sigma_{\rm{j}}$ are pressure and cross-section of the jet. 
The gas flowing into the jet head through the 
shocks would then flows laterally to the 
cocoon, transferring energy from the head. 
Assuming the radiation loss is negligible since 
the AGN disk is highly opaque (see Appendix 
\ref{Appendix-opacity}), 
the energy deposited into the cocoon is 
$E_{\rm{c}} \simeq L_{\rm{jet}}(t-z_{\rm{h}}/c)$, 
where $t$ is the duration of the 
jet propagation and $z_{\rm{h}}= \beta_{\rm{h}} c t$ 
is the height of the jet head. Besides, 
a high $\beta_{\rm{h}}$ would cause the shock 
and the energy flux being radiation dominated
\citep{Yalinewich19}, so the cocoon's pressure can be 
estimated as
\begin{equation}\label{Pc}
P_{\rm{c}}=\frac{E_{\rm{c}}}{3V_{\rm{c}}}\simeq\frac{L_{\rm{jet}}(1-\beta_{\rm{h}})}{3\pi c^3\beta_{\rm{h}}\beta_{\rm{c}}^2t^2},
\end{equation}
where the cocoon is set as a cylinder with height $z_{\rm{h}}$
and radius $r_{\rm{c}}=\beta_{\rm{c}}t$, of which 
$\beta_{\rm{c}} \sim \sqrt{P_{\rm{c}}/(\rho_{\rm{AGN}}c^2)}$ is its 
lateral expansion velocity. For a high $P_{\rm{c}}$, the 
cocoon would collimate the jet, of which the cross-section 
is constrained to
\begin{equation}\label{sigma}
\Sigma_{\rm{j}}\simeq\frac{L_{\rm{jet}}\theta_0^2}{4cP_{\rm{c}}},
\end{equation}
where $\theta_0$ is the initial opening angle of the jet,
which is set by a fiducial value of $0.17 ~(10^\circ)$.
Combining Equation (\ref{vh})-(\ref{sigma}), $\tilde{L}$ 
can be expressed by
\begin{equation}\label{Lcol2}
	\tilde L = \left(\frac{L_{\rm{jet}}}
	{\rho_{\rm{AGN}}t^2\theta_0^4c^5}\right)^{2/5}.
\end{equation}

\begin{figure*}
	\begin{center}
		\includegraphics[width=0.4\textwidth]{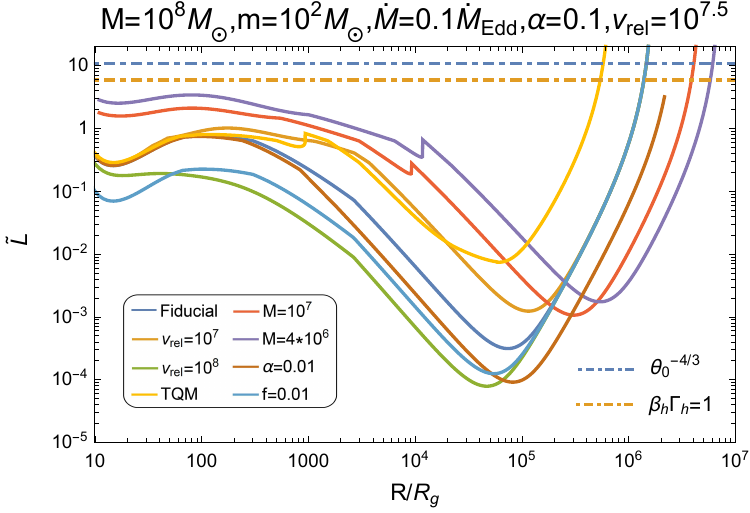}
		\includegraphics[width=0.4\textwidth]{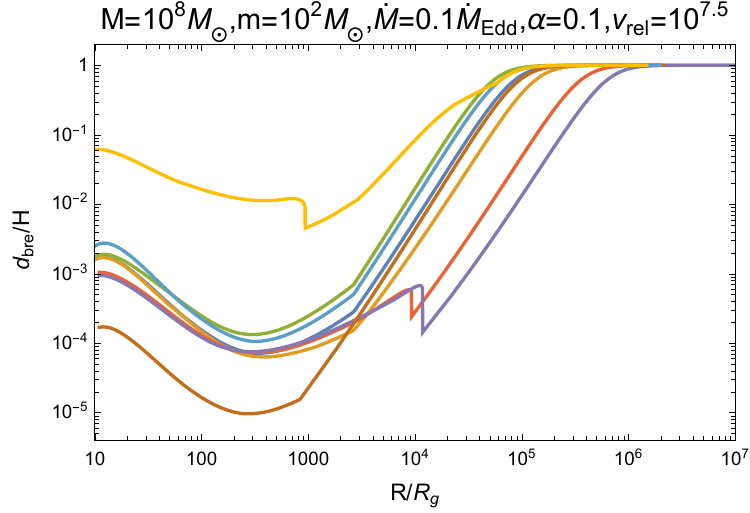}
		\includegraphics[width=0.4\textwidth]{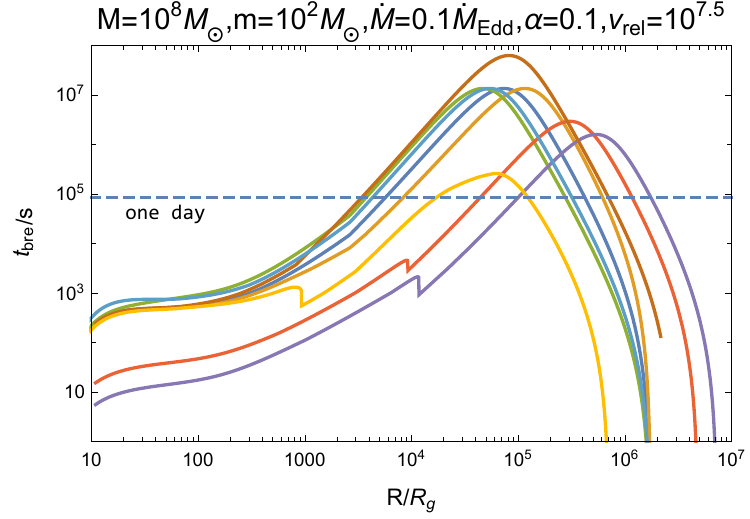}
	\end{center}
	\caption{Various properties of the jet at its breakout, consisting of 
		the jet collimation parameter $\tilde{L}$, the thickness between 
		the jet head and the AGN disk surface $d_{\rm{bre}}$, and the time of 
		jet propagation before the breakout $t_{\rm{bre}}$. The 
		blue and orange dot-dash lines in the $\tilde{L}$ panel severally 
		represents the criterion of collimation, of which the jet is 
		uncollimated when $\tilde{L}>\theta_0^{-4/3}$ \citep{Bromberg11}, and the 
		head's motion being relativistic. The blue dashed line in the
		$t_{\rm{bre}}$ panel represents the time equaling to one day.}
	\label{Fig:Bre}
\end{figure*}

As the jet approaches the AGN disk surface, photons behind the 
shock would diffuse faster than the jet head propagation, 
leading to the shock breakout. We set the distance between 
the jet head and the disk surface at breakout as 
$d_{\rm{bre}}$, where the photon diffusion time, 
$t_{\rm{diff,h}}\sim d_{\rm{bre}}^2\kappa_{\rm{AGN}}\rho_{\rm{AGN}}/c$,
equals to the head dynamical timescale 
$t_{\rm{dyn}}\sim d_{\rm{bre}}/\beta_{\rm{h}}c$, i.e.,
\begin{equation}\label{dbre}
	d_{\rm{bre}}=\frac{1}{\kappa_{\rm{AGN}}\rho_{\rm{AGN}}\beta_{\rm{h}}},
\end{equation}
where we assume $\kappa_{\rm{AGN}}=0.34\cm^2\g^{-1}$ for 
the Thompson scattering of ionized gas (a detailed discussion 
is shown in Appendix \ref{Appendix-opacity}). We
calculate $\tilde{L}$ (i.e. $\beta_{\rm{h}}$) at the time of breakout
$t_{\rm{bre}}=(H-d_{\rm{bre}})/\beta_{\rm{h}}c$ 
via Equation (\ref{vh}),
(\ref{Lcol2}) and (\ref{dbre}), where we ignore the case of the 
AGN disk being initially optically thin, i.e., 
$\kappa_{\rm{AGN}}\rho_{\rm{AGN}} H<1$; the results are shown in 
Figure \ref{Fig:Bre}. First, for various system parameters, 
almost all jets are Newtonian and possess 
$\tilde{L}<\theta_0^{-4/3}$, thereby are collimated by 
cocoons \citep{Bromberg11}; also, the head velocities are 
large enough to generate radiation-dominated shocks \citep{Yalinewich19}. 
So, our calculations under the premise of collimation and 
radiation-domination are valid. Second, except for the 
outer AGN disk regions, the thickness $d_{\rm{bre}}\ll H$, indicating 
that the jets adequately interact with the environment and the 
radiation just starts to escape near the disk surface. 
Conversely, in regions with large $R$, the diffusion
of photons from the shock at early stage can
significantly impact the propagation of jets; 
fortunately, the jet's power is weak at these radii,
as shown in Figure \ref{Fig:Ljet}, 
allowing us to disregard the resulting inaccuracies. 
Third, the values of $t_{\rm{bre}}$ are distributed over 
a large range, depending mainly on the properties of  
AGN disk rather than the kick velocity $v_{\rm{k}}$ and the 
feedback efficiency $f$.

In addition, we can derive the cocoon's properties at 
the jet breakout, i.e., the total energy in the cocoon 
\begin{equation}\label{Ec}
	E_{\rm{c}} \simeq L_{\rm{jet}} t_{\rm{bre}} (1-\beta_{\rm{h}}(t_{\rm{bre}})),
\end{equation}
and the volume of the cocoon
\begin{equation}\label{Vc}
	V_{\rm{c}} \simeq \pi \beta_{\rm{h}}(t_{\rm{bre}}) \beta_{\rm{c}}(t_{\rm{bre}})^2 c^3 t_{\rm{bre}}^3.
\end{equation}
At the jet breakout, radiation losses at the head shock 
come into play, impeding effective energy transfer 
to the cocoon. Subsequently, the cocoon undergoes 
independent evolution driven by its initial energy 
and volume of Equation (\ref{Ec}) and (\ref{Vc}).

\section{EM emission powered by a kicked remnant-driven jet}
\label{section4}
During the jet breakout phase with 
$t_{\rm{diff,h}}<t_{\rm{dyn}}$,
photons would escape from the head shock, thereby
having the potential to produce detectable emission.
Besides, during the propagation within the AGN disk, 
the jet deposits a huge amount of energy to the 
cocoon, which would thus be a competent emission 
source as well. The cocoon comprises two components 
(see panel (a) of Figure \ref{Fig:sketch}): 
the so-called disk and jet cocoon, composed of the 
AGN disk medium and the jet material crossing the 
head's shock, respectively, separated by a contact 
discontinuity \citep[e.g.][]{Bromberg11}. Both components 
are expected to generate emission during their 
expansion following breakout 
(see panel (b) of Figure \ref{Fig:sketch}). 

Therefore, in what follows, we investigate 
the emission during jet breakout, 
as well as the cooling emission from both 
disk cocoon and jet cocoon, focusing 
on the observed properties, comprising luminosity, 
duration, and temperature, of the emission. 

\begin{figure*}
	\begin{center}
		\includegraphics[width=0.32\textwidth]{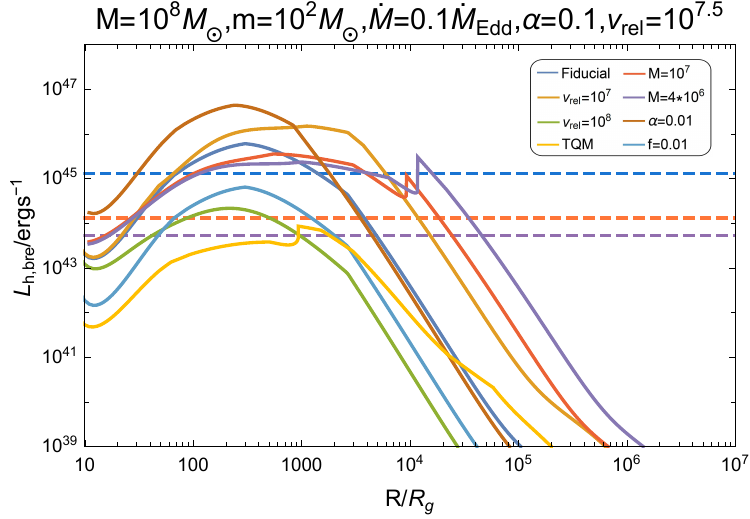}
		\includegraphics[width=0.32\textwidth]{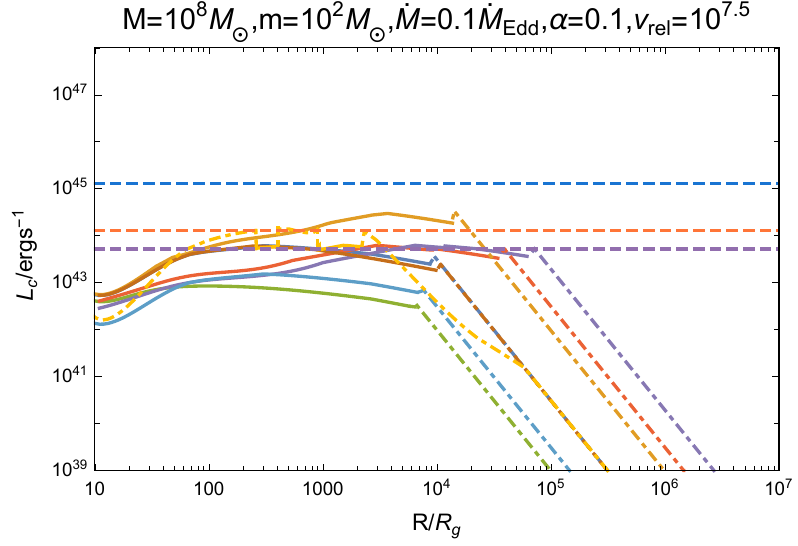}
		\includegraphics[width=0.32\textwidth]{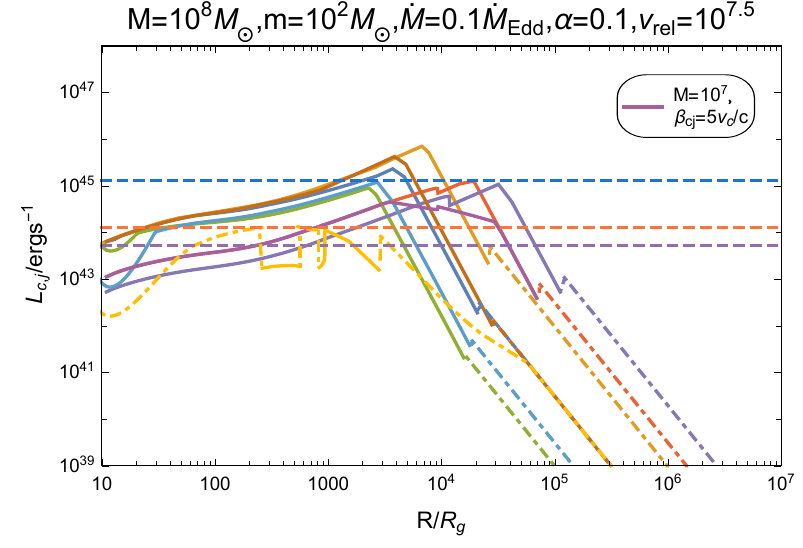}
		\includegraphics[width=0.32\textwidth]{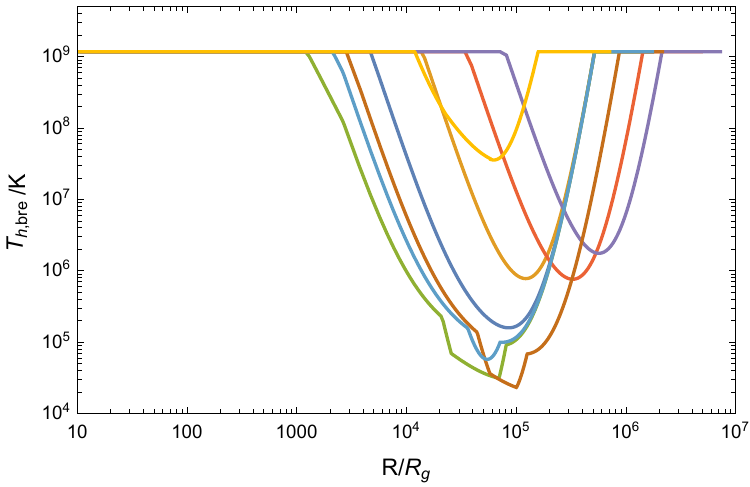}
		\includegraphics[width=0.32\textwidth]{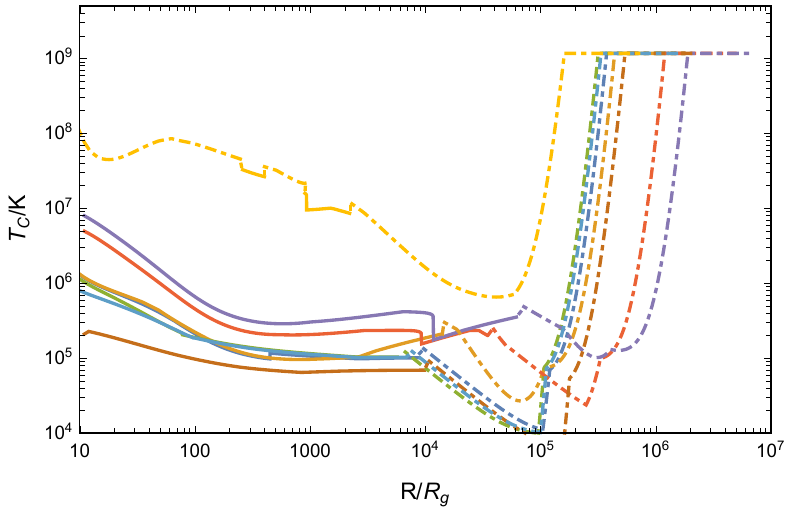}
		\includegraphics[width=0.32\textwidth]{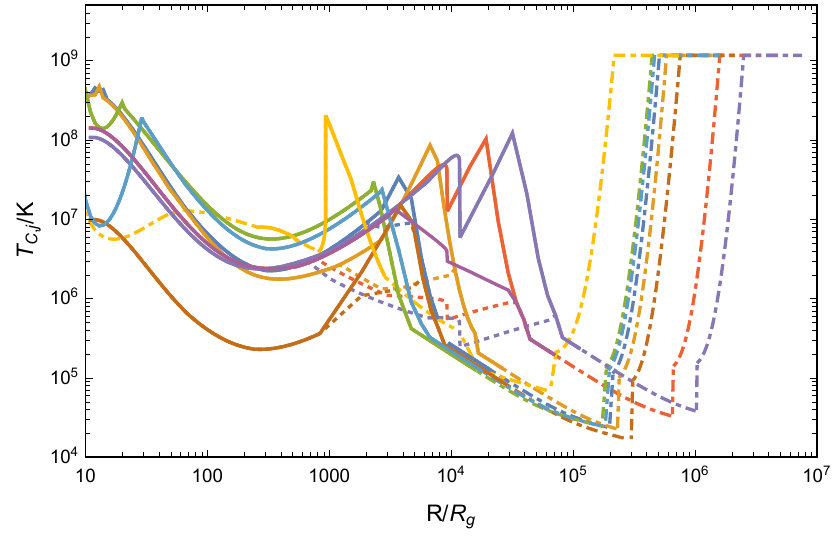}
		\includegraphics[width=0.32\textwidth]{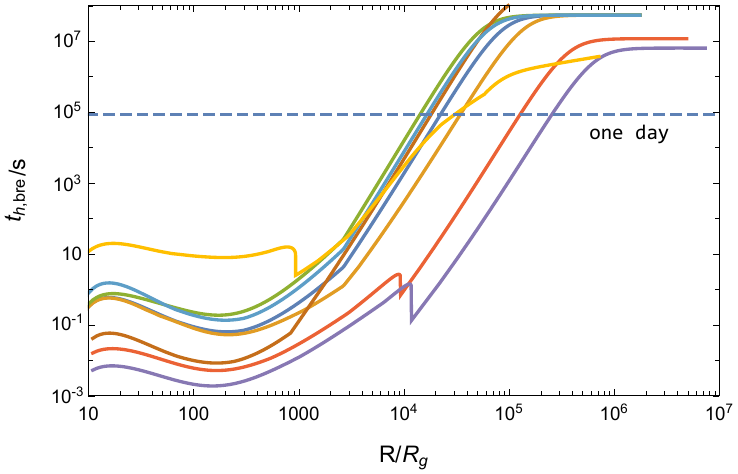}
		\includegraphics[width=0.32\textwidth]{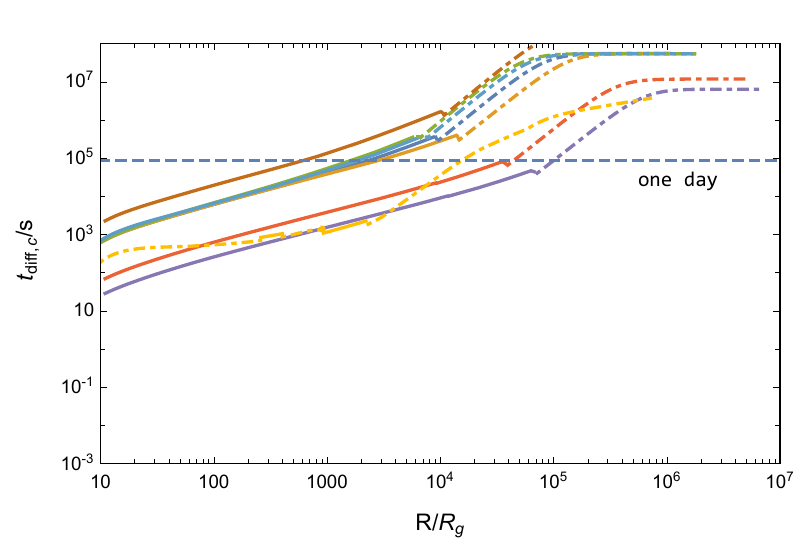}
		\includegraphics[width=0.32\textwidth]{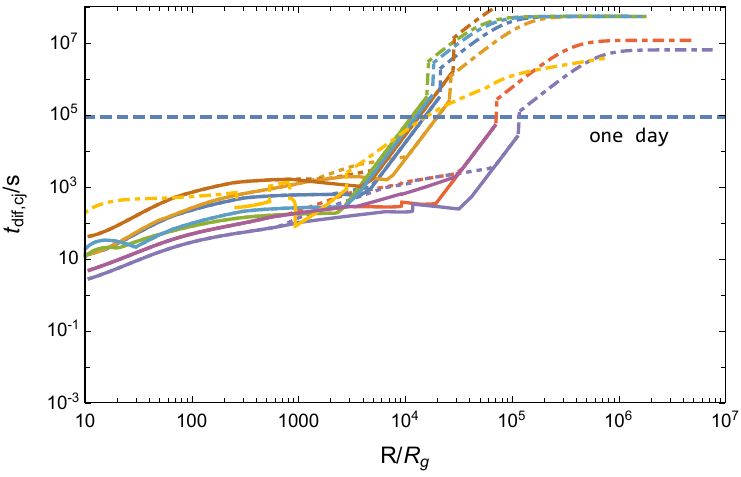}
		\includegraphics[width=0.32\textwidth]{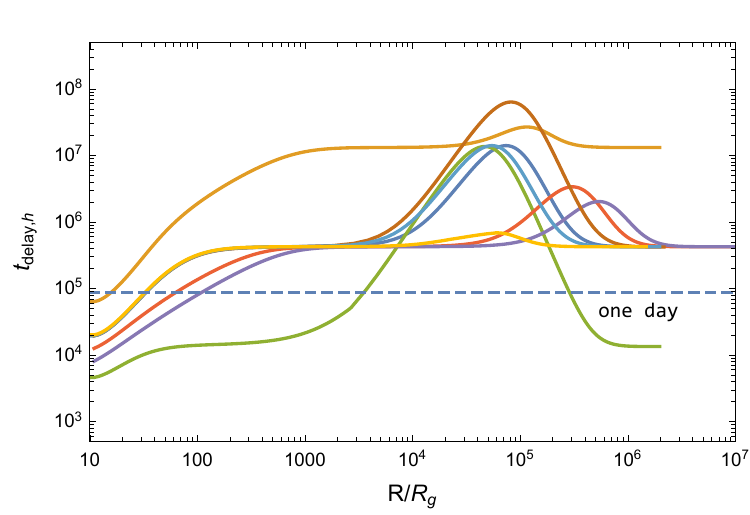}
        \includegraphics[width=0.32\textwidth]{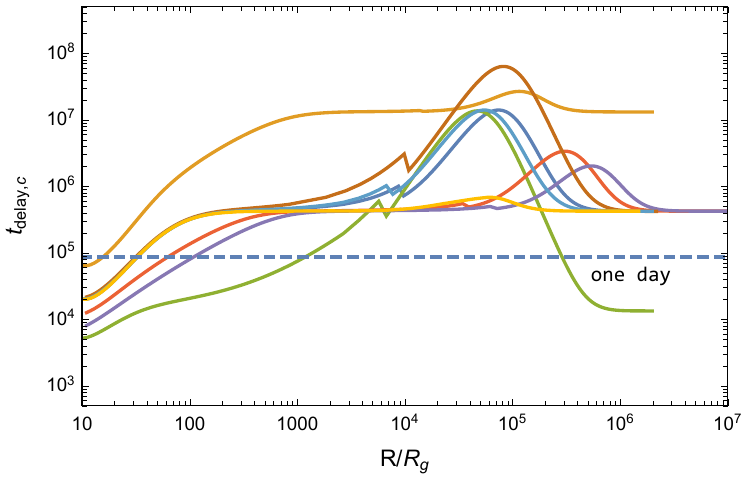}
        \includegraphics[width=0.32\textwidth]{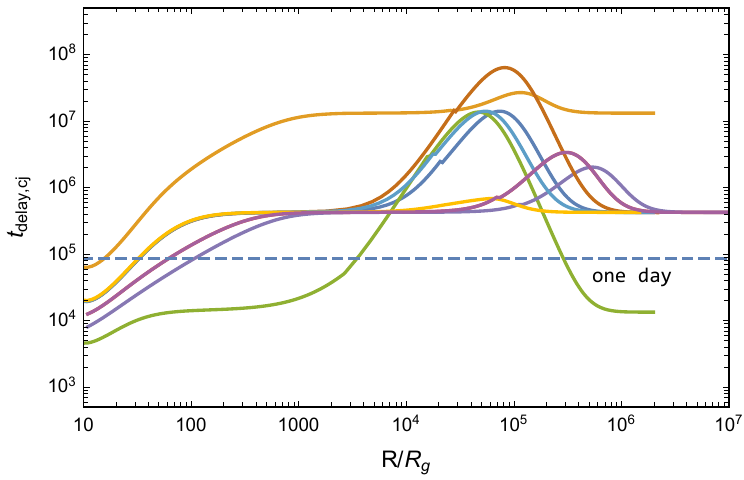}
	\end{center}
	\caption{Observed properties of the emission during 
		jet breakout (the first column), from cooling 
		disk cocoon (the second column), and cooling jet 
		cocoon (the third column), including luminosity, 
		temperature, duration timescale, and time delay. 
		The blue, orange, and purple dashed 
		lines in the first row represent the bolometric 
		luminosity of the AGN disk 
		$L_{\rm{bol,AGN}}=0.1 \dot{M} c^2$. 
	    The blue dashed lines in the third 
	    and forth row represent the 
	    time equaling to one day. The dot dash lines in the second 
        and third columns show the $R$ regions where photons can 
        escape from the cocoons before the jet breakout, and thereby 
        the jet evolution is inaccurate. In addition, in panels of 
        $T_{\rm{cj}}$ and $t_{\rm{diff,cj}}$, the dotted lines 
        represent the properties of emission with 
        $L_{\rm{cj}}=L_{\rm{bol,AGN}}$, i.e., nearly equaling to the 
        AGN disk background radiation. In the third column we 
        additionally consider the case with parameters 
        $M=10^{7}\rm{M}_{\odot}$ and $\beta_{\rm{cj}}=5v_{\rm{c}}/c$, 
        indicating the cooling emission of a jet cocoon, which
        is specifically mixed with the disk cocoon.}
	\label{Fig:rad}
\end{figure*}

\subsection{Emission during jet breakout}
During jet breakout, the head continuously shocks the AGN 
disk medium, but at the moment, photons behind the 
radiation-dominant shock can diffuse out and take away 
most of the energy. Thus, the luminosity of the emission 
can be estimated as \citep[e.g.][]{Tagawa23a}
\begin{equation}
	L_{\rm{h,bre}} \simeq \Sigma_{\rm{j}} \rho_{\rm{AGN}} \beta_{\rm{h}}^3 c^3,
\end{equation}
and the duration can be estimated as the diffusion timescale,
\begin{equation}
t_{\rm{h,bre}} \simeq t_{\rm{diff,h}} = 
\frac{1}{\kappa_{\rm{AGN}}\rho_{\rm{AGN}}\beta_{\rm{h}}^2c}.
\end{equation}

We focus on the thermal emission. The thermal equilibrium 
temperature behind the shock can be calculated via the jumping 
condition for gas with the adiabatic index $\gamma=4/3$ 
\citep[e.g.][]{Zhang18}, i.e.,
\begin{equation}
	T_{\rm{BB,hbre}} = 
	\left(\frac{18}{7a}\rho_{\rm{AGN}}\beta_{\rm{h}}^2c^2\right)^{0.25},
\end{equation}
where $a$ is the radiation constant. However, because the 
jet head's velocity is high, as shown in Figure \ref{Fig:Bre}, 
radiation behind the shock may deviate from thermal equilibrium 
if an insufficient number of photons are generated by
gas via free-free emission. In such a case, the shock would attain 
a thermal Compton equilibrium at 
temperature $>T_{\rm{BB,hbre}}$. Moreover, 
for even higher temperatures, pair production processes 
will be invoked to restrict the temperature 
\citep[e.g.][]{Budnik10, Katz10, Nakar10, Nakar12}.

To determine the radiation temperature, we 
employ the methodology described in \cite{Nakar10}. 
A thermal coupling coefficient can be utilized 
as an indicator of thermal equilibrium, i.e.,
\begin{equation}
\eta \equiv \frac{n_{\rm{BB}}\left(T_{\rm{BB,hbre}}\right)}
{t_{\rm{h,bre}} \dot{n}_{\rm{ph},\rm{ff}}\left(T_{\rm{BB,hbre}}\right)}, 
\end{equation}
where $n_{\rm{BB}}\sim aT_{\rm{BB}}^4/3kT_{\rm{BB}}$ 
is the photon number density required for 
thermal equilibrium, $k$ is the Boltzmann constant, 
and $\dot{n}_{\rm{ph},\rm{ff}} \approx 
3.5\times 10^{36}\s^{-1}\cm^{-3} \rho_{\rm{AGN}}^2 T^{-1/2}$.
For $\eta<1$, the radiation can achieve thermal equilibrium 
with temperature $T_{\rm{BB,hbre}}$. For $\eta>1$, 
photons are starved to reach the blackbody, and electrons
would be in Compton equilibrium with radiation at 
$T \xi(T)^2=T_{\rm{BB}}\eta^2$, of which the 
Comptonization correction factor is
\begin{equation}
	\xi(T) \approx \max\left\{1,\frac{1}{2}\ln\left(y_{max}\right)
	\left[1.6+\ln\left(y_{max}\right)\right]\right\},
\end{equation}
where $y_{max}=3~(\rho/10^{-9}\g\cm^{-3})^{-1/2}
(T/100\eV)^{\rm{9/4}}$. Moreover, pair 
production process limits the highest temperature, 
for which we set $\sim 100~\rm{keV}$ \citep[e.g.][]{Nakar17}.
Combining these analyses, the radiation temperature 
during the jet breakout can be determined by
\begin{equation}\label{Thbre}
	T_{\rm{h,bre}}=\min\left\{T_{\rm{h,Comp}}, 100~\rm{keV} \right\},
\end{equation}
where the Comptonization modified temperature is
\begin{equation}
T_{\rm{h,Comp}}=T_{\rm{BB,hbre}}
\begin{cases}
\eta^2/\xi(T_{\rm{h,Comp}})^2 & \eta>1  \\
1  & \eta<1.
\end{cases}
\end{equation}

The observed properties of the emission are shown 
in the first column of Figure \ref{Fig:rad}. There 
exists a wide range of $R$ that satisfies 
$L_{\rm{h,bre}}>L_{\rm{bol,AGN}}$, with $T_{\rm{h,bre}}$
generally exceeding $O(10^6)\rm{K}$ at these regions, and 
consequently, the breakout emission can outshine the AGN disk 
background in X-ray bands. However, the observed emission 
typically exhibits a short duration with 
$t_{\rm{h,bre}}<O(10^2)\s$, which is significantly shortened
for a lighter AGN case, making the flare slightly hard 
to be detected. In addition, changing the system parameters, 
we find that the observable regions enlarge for the 
case with a smaller SMBH mass $M$ or a
lower kick velocity $v_{\rm{k}}$, and the emission becomes 
more intense in the case of a smaller $\alpha$ or 
a lower $v_{\rm{k}}$, all of which facilitate the 
observation.

\subsection{Emission from cooling disk cocoon}

Given the resemblance, we investigate the 
emission produced by the expanding cocoon 
through an analogy with the cocoon's radiation 
subsequent to the breakout of a gamma-ray burst 
jet from the star \citep{Nakar17}.

The energy deposited in the disk cocoon is 
$E_{\rm{cd}} \approx E_{\rm{c}}/2$, primarily 
in the form of radiation. 
Same as the jet, the cocoon also undergoes a 
breakout from the AGN disk. 
For an opaque cocoon, following the breakout,
the initial radiation energy is converted into 
the gas kinetic energy to drive its expansion. 
The terminal velocity of the  
disk cocoon material with mass, 
$m_{\rm{c}}\simeq \rho_{\rm{AGN}} V_{\rm{c}}$, 
can be estimated as 
$v_{\rm{c}}\simeq \sqrt{2 E_{\rm{cd}}/m_{\rm{c}}}$.
Simplifying the expanding cocoon as a spherical 
shell with radius 
$r_{\rm{c}}=V_{\rm{c}}^{1/3}+v_{\rm{c}}t_{\rm{exp}}$ at 
time $t_{\rm{exp}}$ after the breakout, photons  
would diffuse out from the shell when its optical depth, 
$\tau_{\rm{c}} \simeq \kappa m_{\rm{c}}/4 \pi r_{\rm{c}}^2$, 
roughly equals to $c/v_{\rm{c}}$, the diffusion 
timescale is then given by
\begin{equation}\label{tdiffc}
	t_{\rm{diff,c}} \simeq \tau_{\rm{c}} \frac{r_{\rm{c}}}{c}
	=\left(\frac{\kappa m_{\rm{c}} }{4\pi v_{\rm{c}} c}\right)^{1/2},
\end{equation}
where $\kappa=0.34$ for the ionized gas. Prior to 
photon effective diffusion, the cocoon undergoes adiabatic 
cooling, resulting in a reduction of internal 
radiation energy to
$E_{\rm{r,c}}\sim E_{\rm{cd}}V_{\rm{c}}^{\frac{1}{3}}/r_{\rm{c}}$, 
which is subsequently released over $t_{\rm{diff,c}}$. 
Thereby, the luminosity of the disk cocoon cooling emission 
is
\begin{equation}\label{Lc}
	L_{\rm{c}}=\frac{E_{\rm{r,c}}}{t_{\rm{diff,c}}}
	\simeq\frac{2\pi c E_{\rm{c}}V_{\rm{c}}^{1/3}}{\kappa m_{\rm{c}}}.
\end{equation}

The temperature of the disk cocoon at the 
breakout, $T_{\rm{c,bre}}$, can be estimated 
similarly to Equation (\ref{Thbre}), where the 
thermal equilibrium temperature is modified to
\begin{equation}
	T_{\rm{BB,c}} = 
	\left(\frac{E_{\rm{c}}}{2 a V_{\rm{c}}}\right)^{0.25},
\end{equation}
and the thermal coupling  coefficient is changed to
\begin{equation}
\eta_{\rm{c}} \equiv \frac{n_{\rm{BB}}\left(T_{\rm{BB,c}}\right)}
{t_{\rm{bre}} \dot{n}_{\rm{ph},\rm{ff}}\left(T_{\rm{BB,c}}\right)} . 
\end{equation}
After adiabatic cooling of radiation
during the cocoon expansion, the temperature becomes
\begin{equation}\label{Tc}
	T_{\rm{c}}=T_{\rm{c,bre}}\frac{V_{\rm{c}}^{1/3}}{r_{\rm{c}}},
\end{equation}
which is, namely, the temperature of the disk cocoon 
cooling emission.

The above calculations are based on the assumption that 
$\tau_{\rm{c}} \gg c/v_{\rm{c}}$ at the jet breakout, 
indicating that photons initially remain trapped
within the disk cocoon. Conversely, 
when $\tau_{\rm{c,bre}}\sim 
\kappa m_{\rm{c}}/4 \pi V_{\rm{c}}^{2/3} < c/v_{\rm{c}}$, 
photons would escape from 
the cocoon during the jet propagation within the AGN disk. 
In such a case, the failure of assuming an adiabatic 
cocoon leads to inaccuracies in the calculation of 
jet propagation and collimation 
in Section \ref{sec-jet-propagation}, 
and the intricate cooling process should be 
considered to investigate the concrete evolution of 
both jet and cocoon. For completeness, keeping the 
inaccuracy in mind, we estimate the observed 
properties of the disk cocoon cooling 
emission as follows: when 
$\tau_{\rm{c,bre}} < c/v_{\rm{c}}$, 
the luminosity of emission 
is set as the jet power, 
$L_{\rm{c}} \simeq 0.5 L_{\rm{jet}}$,
i.e., photons escape after 
being spilled from the jet head;
the duration of emission is set as
$t_{\rm{c}} \simeq t_{\rm{bre}} + t_{\rm{h,bre}} $,
i.e., the duration of the jet releasing energy 
into the disk plus the time of radiation diffusing 
out from the disk. And the temperature is set as 
$T_{\rm{c,bre}}$ at the jet breakout.

The observed properties of the  
emission are shown in the second column 
of Figure \ref{Fig:rad}, where 
the dot dash lines indicate $R$ regions with
$\tau_{\rm{c,bre}} < c/v_{\rm{c}}$, for which 
the radiation properties are inaccurate. 
First, from the perspective of total 
luminosity, $L_{\rm{c}}$ is generally lower than
$L_{\rm{bol,AGN}}$, rendering the observation of 
the disk cocoon cooling emission challenging.
Second, the radiation temperatures 
predominantly fall within UV bands except for 
the TQM model case, of which 
the temperature is higher because of the lower 
disk density resulting in an inefficient thermalization.
Furthermore, the temperature primarily relies on the AGN disk 
properties rather than the specific BH accretion processes. 
Third, the emission duration, or the 
radiation diffusion timescale, primarily relies
on the AGN disk properties, and typically spans less 
than one day.  
Additionally, our artificial setup for the emission 
properties in regions where 
$\tau_{\rm{c,bre}} < c/v_{\rm{c}}$  
has a negligible impact on the observation, because, 
in general, $L_{\rm{c}}$ is significantly less than 
$L_{\rm{bol,AGN}}$, thereby, the emission is unable 
to outshine the AGN background radiation and 
remain undetectable.

\subsection{Emission from cooling jet cocoon} 

Similar to the disk cocoon, the internal energy of 
the jet cocoon is radiation dominated with a total 
deposited energy $\sim E_{\rm{c}}/2$, which 
subsequently accelerates the rarer shocked materials 
after the jet breakout.
Numerical simulations show that, during the jet 
propagation within dense medium, partial material 
mixing would take place between the two cocoon 
components \citep[e.g.][]{Morsony07, Nakar17, 
Gottlieb20, Gottlieb21}.
The mixed jet cocoon exhibits stratification, where
the material near the base of jet is more mixed,
thus colder and heavier; conversely, the material closer to
the jet head experiences less mixing, thus hotter and 
lighter \citep{Eisenberg22}. 
Additionally, the energy of the jet cocoon 
exhibits a roughly flat distribution per logarithmic scale of 
terminal proper velocity, corresponding to the energy 
per baryon of the shocked materials 
\citep[e.g.][]{Gottlieb20, Gottlieb21}. 
So we estimate the fraction of the cocoon's 
energy deposited in material with a specific 
$\Gamma_{\rm{cj}} \beta_{\rm{cj}}$ 
as $f_{\Gamma \beta} \sim 0.1$ following \cite{Nakar17}, with 
the Newtonian shocked jet material considered as 
a representative emission source due to its longer 
duration compared to the shorter-lived
emission from the relativistic material. Though, 
in actuality, radiation is continuously emitted early 
from the relativistic material and later from 
the Newtonian material \citep{Nakar17}.

The mass of the jet cocoon material with a terminal velocity 
$\Gamma_{\rm{cj}} \beta_{\rm{cj}} < 1$ $(\beta_{\rm{cj}} \lesssim 0.7)$ 
is estimated as
\begin{equation}
	m_{\rm{cj}} \simeq \frac{f_{\Gamma \beta} E_{\rm{c}}}{\beta_{\rm{cj}}^2 c^2},
\end{equation}
and the volume of the jet cocoon at the breakout is
\begin{equation}
	V_{\rm{cj}} \simeq \Sigma_{\rm{j}} (H-d_{\rm{bre}}).
\end{equation}
We define a critical velocity $\beta_{\rm{cj,cr}}$ 
to represent the case of which photons begin to escape 
from the jet cocoon roughly at the jet breakout, i.e., 
$\tau_{\rm{cj,bre}} \simeq \kappa m_{\rm{cj}}/V_{\rm{cj}}^{2/3} =1/\beta_{\rm{cj,cr}}$,
and then
\begin{equation}
	\beta_{\rm{cj,cr}}= \frac{\kappa f_{\Gamma \beta} E_{\rm{c}}}
	{V_{\rm{cj}}^{2/3} c^2},
\end{equation}
where still we set $\kappa = 0.34$ for an ionized medium.
The photon is coupled with the jet cocoon material with
$\beta_{\rm{cj}}<\beta_{\rm{cj,cr}}$ at the jet breakout. 
Subsequently, the cocoon with 
mass $m_{\rm{cj}}$ undergoes expansion and 
adiabatic cooling before eventually allowing for 
photon escape. This process remains largely unaffected 
by the faster material with $\beta_{\rm{cj}}>\beta_{\rm{cj,cr}}$, 
due to the stratified jet cocoon exhibits preferential 
escape of the upper hotter material followed 
by the lower colder material \citep{Eisenberg22}. 

Similar to Equation (\ref{Lc}), the luminosity of the 
jet cocoon cooling emission is estimated as
\begin{equation}\label{Lcj}
	L_{\rm{cj}}
	=\frac{2\pi c f_{\Gamma \beta} E_{\rm{c}}V_{\rm{cj}}^{1/3}}
	{\kappa m_{\rm{cj}}}
	=\frac{2 \pi \beta_{\rm{cj}}^2 c^3 V_{\rm{cj}}^{1/3}}{\kappa},
\end{equation}
which is in proportion to $\beta_{\rm{cj}}^2$. Considering 
only the Newtonian part of the shocked jet cocoon, we set
$\beta_{\rm{cj}}=0.7$ for cases with 
$\beta_{\rm{cj,cr}}>0.7$, 
and set $\beta_{\rm{cj}}=\beta_{\rm{cj,cr}}$ for cases with
$0.7>\beta_{\rm{cj,cr}}>v_{\rm{c}}/c$, 
to calculate the peak luminosity. And similar to 
Equation (\ref{tdiffc}), the emission duration is 
estimated as the photon diffusion timescale,
\begin{equation}
	t_{\rm{diff,cj}} \simeq 
	\left(\frac{\kappa m_{\rm{cj}} }{4\pi \beta_{\rm{cj}} c^2}\right)^{1/2}
	= \left(\frac{\kappa f_{\Gamma \beta} E_{\rm{c}} }
	{4\pi \beta_{\rm{cj}}^3 c^4}\right)^{1/2}.
\end{equation}
The radiation temperature $T_{\rm{cj}}$ can be calculated 
similarly as Equation (\ref{Thbre}) and (\ref{Tc}), by 
substituting the initial energy, mass, and volume to 
$f_{\Gamma \beta} E_{\rm{c}}/2$, $m_{\rm{cj}}$, and $V_{\rm{cj}}$.
Moreover, for emission with 
$L_{\rm{cj,peak}}>L_{\rm{bol,AGN}}$, 
we also calculate $t_{\rm{diff,cj}}$ 
and $T_{\rm{cj}}$ at 
$L_{\rm{cj}}=L_{\rm{bol,AGN}}$, 
adopting $f_{\Gamma \beta} = 0.1$, 
to estimate the maximum 
duration for observation and the temperature evolution,
since the energy and mass distribution 
of the jet cocoon, which can be described as 
$dE/dv\propto v^{-n}$ and $m(>v)\propto v^{-(n+1)}$ 
\citep[e.g.][]{Nakar17,Piro18}, result in a luminosity 
evolution after the peak following 
$L_{\rm{cj}} \propto t^{-4/(n+2)}$.

In the above we focus on the opaque part of the 
partial mixing jet cocoon, but when
$\beta_{\rm{cj,cr}}<v_{\rm{c}}/c$, the entire jet cocoon 
is transparent even though being fully mixed with 
the disk cocoon to possess a minimum velocity 
$v_{\rm{c}}$. In these cases, in rough, 
we artificially set the luminosity,
the emission duration, and the temperature as
$L_{\rm{cj}}=0.5L_{\rm{jet}}$, 
$t_{\rm{cj}} \simeq t_{\rm{h,bre}} + t_{\rm{bre}}$,
and $T_{\rm{cj,bre}}$ for the 
material with velocity $v_{\rm{c}}$.

The observed properties of the emission are shown 
in the third column of Figure \ref{Fig:rad}. 
First, $L_{\rm{cj}}$ surpasses $L_{\rm{bol,AGN}}$ 
across a wide range of $R$, thereby rendering the 
emission potentially detectable.
The observation of systems with lower mass AGNs 
or slower kicked remnant BHs is more promising, 
owing to the dimmer AGN background radiation or 
the augmented jet power. Second, the 
temperatures of emission typically range 
$O(10^6)-O(10^8)\rm{K}$, hence being observable in 
X-ray bands. Third, the duration of 
emission is generally less than one day, which 
is comparatively shorter than the disk 
cocoon cooling emission due to the higher 
velocity and lower mass of the jet 
cocoon material. Fourth, in the case of emission 
outshining the AGN background, since the energy 
distribution brings about radiation 
evolution, we find that at $L_{\rm{cj}}=L_{\rm{bol,AGN}}$
(dotted lines in panels of $T_{\rm{cj}}$ 
and $t_{\rm{diff,cj}}$ in Figure \ref{Fig:rad}),
whereafter the source tends to become fainter than 
the AGN, the total duration of the observed emission
ranges from a few $O(10^2)\rm{s}$ to $O(10^3)\rm{s}$, 
accompanied by a decrease in temperature 
making for softer radiation.

Note that, in this subsection, we explore the 
jet cocoon cooling emission in analogy with the 
gamma-ray burst jet breaking a collapsing star, 
of which the breakout time, $t_{\rm{bre}}\lesssim 
10\s$ \citep[e.g.][]{Nakar17}, is much less than that of 
a jet breaking the AGN disk studied here. The 
prolongation of the propagation time would 
result in a more efficient mixing 
between the two cocoon components, thereby the 
maximum velocity of the jet cocoon material may be
less than $0.7c$ as set in gamma-ray burst jet system, 
causing a significant decline of the emission 
luminosity due to 
$L_{\rm{cj}} \propto \beta_{\rm{cj}}^2$ demonstrated 
by Equation (\ref{Lcj}). To investigate the 
impact of a reduced $\beta_{\rm{cj}}$, 
we consider a case of efficient mixing, 
where only $10\%$ of the jet cocoon energy 
is deposited into a mixed Newtonian material with 
the maximum velocity 
$\beta_{\rm{cj}}=5v_{\rm{c}}/c$, the relevant
properties of the emission are shown in the 
third column of Figure \ref{Fig:rad}. We find 
that the emission is still observable though 
the mixing leading to a weaker radiation 
with lower temperature. 

\subsection{Time delay between BBH merger and EM emission}
\label{4.4}
Given that the EM radiation arises from the jet triggered 
by the kicked remnant BH, the time delay between a BBH
merger event and its associated EM emission can be 
attributed to four factors: the potential remnant kick time 
$t_{\rm{kick}}$, the jet formation time, 
which is approximately a few $t_{\rm{BHL}}$ \citep{Kaaz23b}, 
the jet breakout time, denoted as $t_{\rm{bre}}$, 
and the time before photons diffusing out effectively, 
taken as $t_{\rm{diff,c}}$ and $t_{\rm{diff,cj}}$, 
for the disk and jet cocoon cooling emission. 
Therefore, the time delay is estimated by
\begin{equation}
	t_{\rm{delay}} \simeq (t_{\rm{kick}}) + 
	t_{\rm{BHL}} + t_{\rm{bre}} + t_{\rm{diff}}, 
\end{equation}
of which the values are shown in the last 
row of Figure \ref{Fig:rad}, where we ignore
$t_{\rm{kick}}$ (values are shown in Figure 
\ref{Fig:abin}). The dominant 
term of $t_{\rm{delay}}$ for all of the three 
emission components shifts from $t_{\rm{BHL}}$ 
to $t_{\rm{bre}}$ as $R$ increases, 
while the contribution from $t_{\rm{diff}}$ 
of the disk and jet cocoon 
is found to be negligible. Moreover, 
in regions of $L>L_{\rm{AGN,bol}}$, 
where the emission would 
outshine the AGN background,
$t_{\rm{delay}} \simeq t_{\rm{BHL}}$ 
is basically satisfied, indicating that 
the time delay between the emergence of 
the observed emission and the BBH merger
is primarily determined by the jet 
formation timescale.
For example, $t_{\rm{delay}}$ is 
approximately $154$, $5$,
and $0.2~days$ for a 
$100M_\odot$ remnant BH with kick velocity  
$10^2$, $10^{2.5}$, and $10^3~\rm{km}\s^{-1}$, 
respectively. Alternatively, considering the 
cavity environment and incorporating $t_{\rm{kick}}$,
$t_{\rm{delay}}$ is about $8.8$, $42.8$, and 
$371.8~days$ for a $100M_\odot$ remnant BH 
kicked with velocity 
$10^{2.5}~\rm{km}\s^{-1}$ at $100$, $10^3$, 
and $10^4R_{\rm{g}}$ of the AGN disk 
with a $10^8M_\odot$ SMBH, respectively.

\section{Discussion}
\label{section5}

\begin{figure*}
	\begin{center}
		\includegraphics[width=0.4\textwidth]{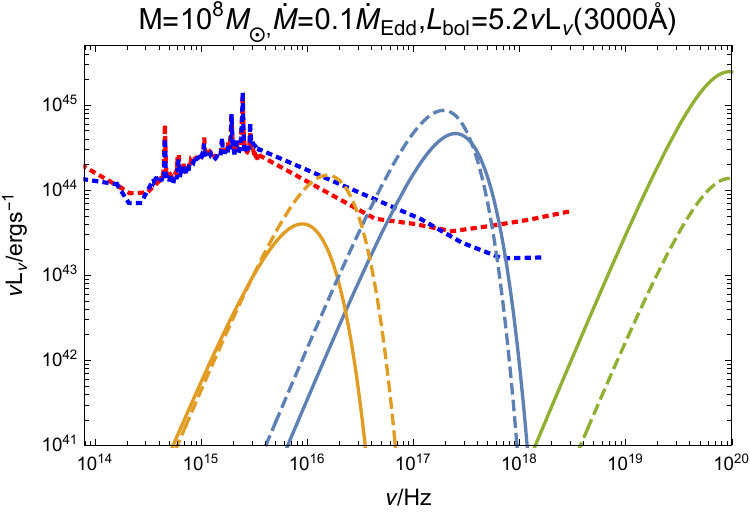}
		\includegraphics[width=0.4\textwidth]{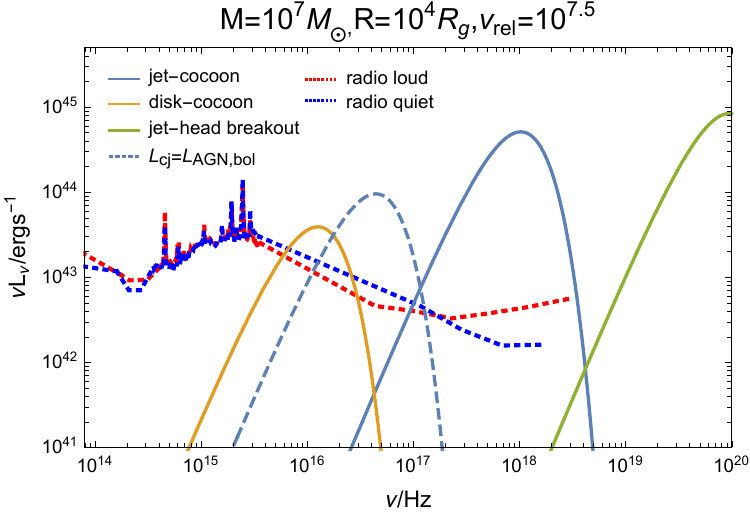}
	\end{center}
	\caption{Spectra for emission from the breakout jet-head  
		(green line), the cooling disk cocoon (golden line), 
		and the cooling jet cocoon (navy-blue line). To 
		express the AGN background, of which the spectra 
		are shown by the red and blue dotted lines, 
		we take the mean spectral 
		energy distribution of radio loud and radio quiet 
		quasars from \cite{Shang11}, with the use of a scaling law 
		$\zeta(3000\mathring{A})=5.2$ in the correlation of 
		$L_{\rm{AGN,bol}}=\zeta\lambda L_{\lambda}$ \citep{Runnoe12}.
		The solid and dashed lines in $M=10^8M_\odot$ 
		panel represent a $100M_\odot$ BBH merger occurring at 
		$700$ and $10^4R_{\rm{g}}$, generating a 
		kicked remnant with velocity 
		$10^{2.5}$ and $10^{2}~\rm{km}\s^{-1}$,
		respectively. The solid line in $M=10^7M_\odot$ 
		panel represents a $100M_\odot$ BBH merger occurring at 
		$10^4R_{\rm{g}}$ with the remnant kicked to 
		$10^{2.5}~\rm{km}\s^{-1}$, and the dashed line 
		shows the spectra of jet cocoon cooling emission 
		at $L_{\rm{cj}}=L_{\rm{bol,AGN}}$.  
		}
		\label{Fig:example}
\end{figure*}

\subsection{Detectability} 
\label{Detectability}

As demonstrated above, BBH mergers occurring within 
the AGN disk have the potential to generate
observable EM counterparts, such as emission during 
the jet breakout, disk cocoon cooling 
emission, and jet cocoon cooling emission, 
which are powerful to outshine 
the AGN background radiation. Meanwhile, the emission 
exhibits significant variations with respect to $R$ 
(see Figure \ref{Fig:rad}), namely, 
the particular locations of BBH mergers,
which are likely at migration traps 
around $\sim 500-700 R_{\rm{g}}$
\citep[e.g.][]{Bellovary16, Yang19a}, 
or alternatively at more distant radii of
$\sim 0.01~\rm{pc}$ for a $4\times10^6M_\odot$ 
SMBH, corresponding to 
$\sim 5\times 10^4 R_{\rm{g}}$ \citep{Tagawa20a}.
Hence, to predict the detectability of
EM counterparts, we give illustrative examples, 
which depict the feasible BBH mergers in the AGN disk
along with a successful emergence of powerful jet,
the emblematic spectra for emission are shown in 
Figure \ref{Fig:example}.
 
As indicated by the three examples, i.e., 
a $100M_\odot$ BBH merger occurring at $700$, $10^4$, 
and $10^4R_{\rm{g}}$ in the AGN disk with a SMBH 
of $10^8$, $10^8$, and $10^7M_\odot$, generating a 
kicked remnant with velocity $10^{2.5}$,
$10^{2}$, and $10^{2.5}~\rm{km}\s^{-1}$: First, 
the jet breakout emission exhibits a peak in 
hard X-ray bands $\sim 100~\rm{keV}$, which could 
be detected by X-ray telescopes such as 
Swift BAT \citep{Barthelmy05} 
and Fermi GBM \citep{Meegan09}
\footnote{Properties of various telescopes, e.g., 
detection bandpass, sensitivity, and field of view, 
are summarized in Table 3 of \cite{Tagawa23a}.}. 
The duration of the emission is $0.21$, $395$, and 
$0.69\s$, respectively, suggesting a preference 
for the observation of mergers with a 
weakly kicked remnant at large radii in the disk 
of heavier SMBH. 
Second, the disk cocoon cooling 
emission peaks in EUV bands, the duration 
of which is $3.0\times10^4$, 
$3.0\times10^5$, $2.2\times10^4\s$, much 
longer than the two other emission components.
But the emission is unlikely to be 
observed because the EUV photons would be 
absorbed by gas and dust in the host galaxy,
meanwhile, the optical/UV radiation is 
greatly covered by the more prominent AGN 
background. Third, the jet cocoon cooling emission 
peaks in soft X-ray band
at $\sim 1-10~\rm{keV}$, of which the duration is 
$574$, $7561$, $374\s$. Moreover, the energy and 
mass distribution of jet cocoon material causes 
an extension of the observable 
duration, e.g., to $1314\s$ for 
the third example (dashed line in the second panel 
of Figure \ref{Fig:example}), 
and an evolution of spectra to 
softer X-ray band, both of which are beneficial 
to observations via soft X-ray telescopes, e.g., 
Chandra \citep{Weisskopf00}, XMM-Newton \citep{Jansen01}, 
Swift XRT \citep{Burrows05}, Einstein Probe \citep{Yuan15}, 
and to the identification of EM counterparts.  
Additionally, the time delay for 
the three examples is approximately $5$, 
$154$, $5~days$ ($31$, $1314$, $80~days$ if the 
kicked remnant is initially in a cavity), 
respectively, disfavoring 
follow-up observations and identification 
of EM counterparts for merger systems which produce 
remnants with low kick velocity.  

To conclude, a BBH merger event occurring within 
the AGN disk can produce a detectable soft X-ray 
counterpart deriving from the cooling jet cocoon, 
which persists for a period $O(10^3)\s$ and 
appears after $O(10)~days$ following the GW trigger. 
Observations prefer to systems leaving 
over BHs with larger kick velocity, which possess 
relatively short time delay; conversely,
though a more weakly kicked remnant BH can 
generate more powerful emission, 
the large time delay makes it difficult to 
be identified as EM counterparts. 
In the future, with a more precise constraint 
on kick velocity through an accurate measurement 
of GW \citep[e.g.][]{Mahapatra23}, 
we will be able to predict the EM counterparts
more accurately, thereby providing better 
guidance for observations.

\subsection{Jet direction, kick direction, and merger height}

During the above calculations, we have adopted 
a typical system, of which the BBH merger 
occurs at the midplane of an AGN disk, 
generating a remnant BH kicked along the 
disk plane which drives a jet perpendicular 
to the disk plane. In this subsection, 
we discuss implications of varying the 
jet direction, the kick direction, and the 
merger height.

The jet driven by the BZ mechanism propagates along 
the BH spin direction, which may be random for 
mergers taking place in the AGN disk \citep{Tagawa20b}. 
For a jet which is inclined at an angle 
$\theta$ with respect to the AGN disk 
angular momentum direction, the propagation 
distance before its breakout is 
extended from $\sim H$ to $\sim H/\cos\theta$.
Compared to the vertical case, at the jet breakout, 
we find approximately
$t_{\rm{bre}} \propto \cos\theta^{-5/3}$, 
$\tilde{L} \propto \cos\theta^{4/3}$, and 
$\beta_{\rm{h}} \propto \cos\theta^{2/3}$, 
indicating that the inclined jet spends more 
time to propagate within the AGN disk, and is 
more significantly collimated with a lower
head's velocity. Also, as 
$E_{\rm{c}}\propto \cos\theta^{-5/3}$, more jet 
energy is deposited into the cocoon. 
Assuming that the jet breaking out from the head
\citep{Tagawa23a}, we investigate the properties 
of associated emission. For jet breakout 
emission, the luminosity would decrease as 
$L_{\rm{h,bre}}\propto \cos\theta^{2/3}$, and 
the duration is extended to 
$t_{\rm{h,bre}}\propto \cos\theta^{-4/3}$. 
The typical radiation temperature undergoes 
a reduction and a more effective thermalization,
with $T_{\rm{BB,hbre}}\propto \cos\theta^{1/3}$ 
and $\eta \propto \cos\theta^{5/2}$. 
The other two emission components exhibit similar 
properties. For disk cocoon cooling emission,
we have $L_{\rm{c}}\propto \cos\theta^{1/3}$, 
$t_{\rm{diff,c}}\propto \cos\theta^{-11/6}$, and
$T_{\rm{BB,c}}\propto \cos\theta^{1/3}$,
$\eta \propto \cos\theta^{17/6}$. Besides, 
radiation temperature after cocoon adiabatic 
cooling varies as
$T_{\rm{c}}\propto T_{\rm{c,bre}} \cos\theta^{1/6}$.
For jet cocoon cooling emission from the material
with a specific $\beta_{\rm{cj}}$, related 
properties are $L_{\rm{cj}}\propto \cos\theta^{-7/9}$, 
$t_{\rm{diff,cj}}\propto \cos\theta^{-5/6}$, and
$T_{\rm{BB,cj}}\propto \cos\theta^{1/6}$,
$\eta \propto \cos\theta^{9/4}$,
$T_{\rm{cj}}\propto T_{\rm{cj,bre}} \cos\theta^{1/18}$,
respectively. In fact, \cite{Rodriguez-Ramirez23}  
investigated a limiting case of a jet propagating 
quasi-parallel to the AGN disk plane with 
$\theta=82^\circ$; an optical flare lasting for
$O(100)~days$ is produced by the disk cocoon, 
of which the temperature is lower and the duration 
is longer than the perpendicular case we studied,
as expected. 

The kick direction of the remnant BH could be 
random because of an isotropic distribution of 
the orbital plane direction of merging BBHs 
(\citealt{Tagawa20b}, but when binary-single interaction 
is 2D and confined roughly to the AGN disk midplane, 
the kick direction is almost along the disk, 
see e.g. \citealt{Samsing22}). For a kick inclined to the 
AGN disk plane, the remnant BH would have moved to a 
high altitude when a jet forms after a 
few $t_{\rm{BHL}}$. Likewise, for the merger of a
BH binary whose orbit around SMBH is misaligned
with the AGN disk, the starting point of the BH-driven 
jet would be located away from the midplane. 
At the side of the AGN disk where the kicked 
BH is closer to the disk surface,
an off-midplane jet would break out of the AGN disk 
more easily with a shorter propagation distance, 
and may generate brighter and shorter-duration 
radiation with higher temperature; on the contrary, 
the same jet needs to propagate a longer distance to 
break out of the other side of the AGN disk, generating 
dimmer and lower-temperature emission with longer duration.

\begin{figure*}
	\begin{center}
		\includegraphics[width=0.5\textwidth]{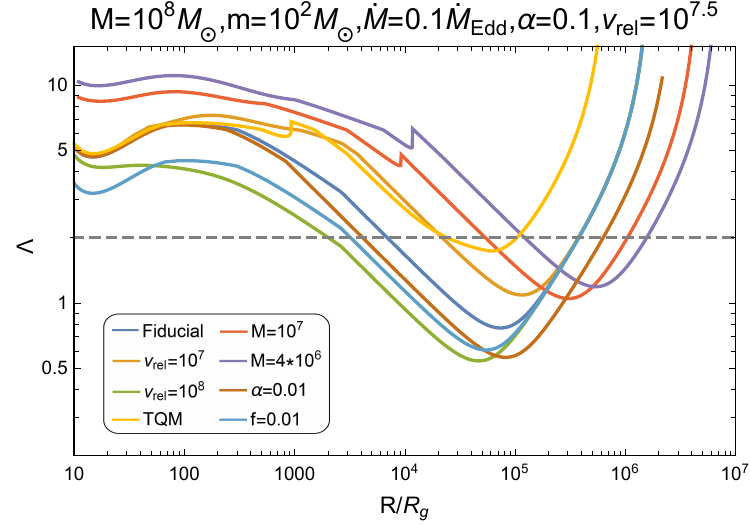}
	\end{center}
	\caption{Criterion $\Lambda$ from Equation 
		(\ref{kink}) as a function of $R$ for 
		different system parameters. The dashed 
		line represents $\Lambda=2$, above 
		which the jet propagation is stable,
		unaffected by the magnetically driven 
		external kink instability.}
		\label{Fig:kink}
\end{figure*}

\subsection{Stability of jet during propagation}

As mentioned above, the jet driven by a kicked 
remnant BH is magnetized. Even though the jet 
behaves as a hot, hydrodynamic one during its 
propagation, the magnetic effects would still 
bring about an external kink instability, 
which generates helical motions of the entire 
jet, increasing its effective cross-section and 
reducing its propagation velocity, even 
causing it to stall \citep{Bromberg16}. We employ a 
criterion proposed by \cite{Bromberg16}, which describes 
a comparison between the instability 
development timescale and the dynamic timescale, 
to examine the instability,
\begin{equation}\label{kink}
	\Lambda \simeq 20\sqrt{\frac{2\pi}{9}}
	\left(\frac{L_{\rm{jet}}}{\rho_{\rm{AGN}} 
	(H-d_{\rm{bre}})^2\gamma_{\rm{j}}^2c^3}\right)^{1/6},
\end{equation}
where $\gamma_{\rm{j}}=\theta_0^{-1}$ is set. 
For $\Lambda>2$, as suggested by \cite{Bromberg16}, 
the external kink is weak and the jet 
propagates similarly to a hydrodynamic one; 
on the contrary, the jet is markedly deformed, 
which would be stalled and fail to break out 
the AGN disk.

Values of $\Lambda$ as a function of $R$ are
shown in Figure \ref{Fig:kink}. $\Lambda$ 
decreases as $R$ increases because of the 
thickening of the AGN disk, and jets are more 
unstable for larger SMBH mass or less jet 
power cases. At relatively large radii, 
where $\Lambda<2$, jets would experience
significant deceleration or disruption 
caused by the external kink instability. 
Consequently, our calculations of the jet 
propagation and the associated EM emission 
may be inaccurate in these regions. However, 
jets with $L_{\rm{jet}}>L_{\rm{AGN,bol}}$, 
which can produce observable emission,  
generally possess $\Lambda>2$, and hence are 
stable to the external kink. Therefore, our 
prediction for the EM counterparts should 
be acceptable.

\section{Summary}
\label{section6}
In this work, we have explored the EM counterparts 
to BBH mergers occurring within the AGN disk.
The merger remnant BH, which undergoes a natal 
kick due to asymmetric GW radiation, would  
traverse the magnetized AGN disk environment and 
launch a jet via the BZ mechanism. Interaction 
between the jet and the AGN disk during its 
propagation leads to jet collimation and energy 
dissipation, thereby driving various EM emissions. 
We have investigated three specific processes 
as sources of radiation, namely, the jet breakout, 
the disk cocoon cooling, and the jet cocoon cooling. 
Among these processes, only the emission from the 
cooling jet cocoon exhibits a higher potential for 
detection. Accordingly, a soft X-ray transient that
outshines the AGN background can be identified as
an EM counterpart, exhibiting a duration of 
$O(10^3)\s$ and emerging after a time delay of 
$O(10)~days$ following the GW trigger. 

We have found that the EM emission properties, 
including luminosity, temperature, duration, 
and time delay, are significantly influenced by 
the GW kick. Therefore, in the future, with improved 
GW measurements, a more precise constraint on kick 
velocity can facilitate the prediction of
EM counterparts to BBH mergers in the AGN disk 
and guide subsequent follow-up 
observations. If such multimessenger signals are
indeed detected, it would confirm the formation 
of binary mergers in the AGN disk, and 
provide valuable insights into the evolution of 
compact objects and binaries embedded in AGN disks.

It should be noted that a key assumption
has been made in this work, namely the
successful launch of a powerful jet by the 
kicked remnant BH, which necessitates a strongly 
magnetized environment. However, the magnetization 
properties of the AGN disk, or even the disk itself, 
have not been well studied. 
For instance, the uncertain degree of ionization 
in the disk outer region may lead to decoupling 
between gas and magnetic field, thereby suppressing 
magnetorotational instability and magnetic field 
amplification, which hinders the jet formation.
So we suggest that a more comprehensive 
understanding of AGN disks is imperative for 
investigating the EM counterparts associated with 
the embedded BBH merger events.

\begin{acknowledgements}
	
We would like to thank the referee for valuable comments and 
helpful suggestions. This work was supported by 
the National SKA Program of China (grant No. 2020SKA0120300), 
and the National Natural Science Foundation of China (grant No. 11833003).	
	
\end{acknowledgements}

\begin{appendix}
\section{Opacity and Optical Depth of AGN Disk in Propagation Path of jet}
\label{Appendix-opacity}

\begin{figure*}
	\begin{center}
		\includegraphics[width=0.45\textwidth]{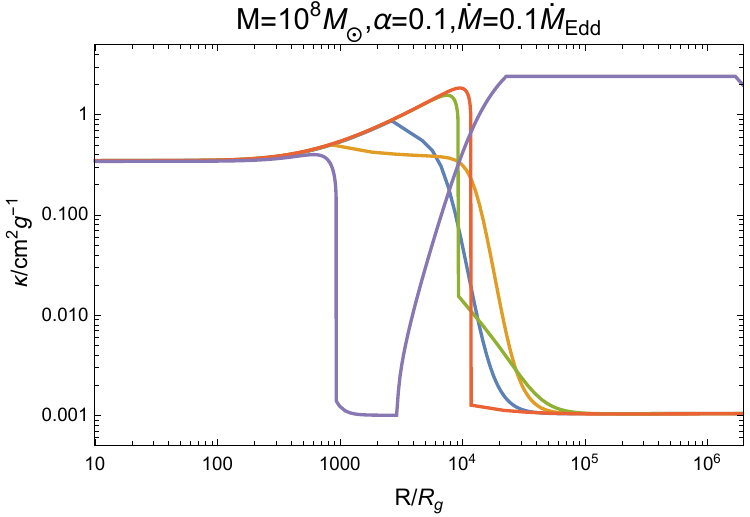}
		\includegraphics[width=0.45\textwidth]{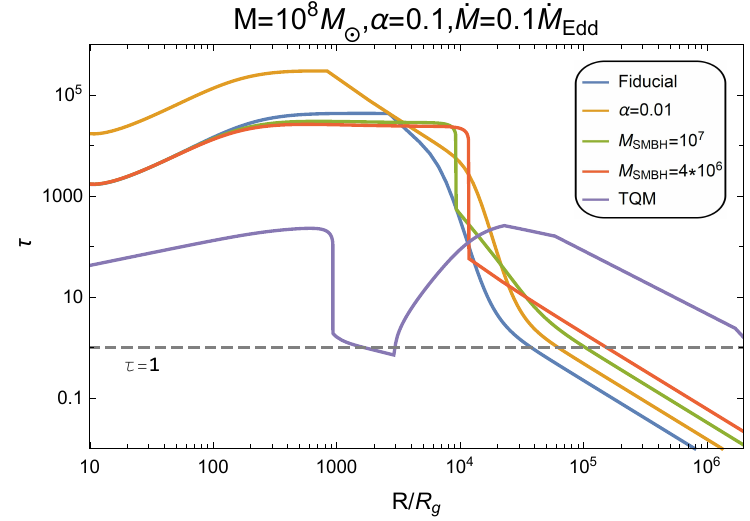}
	\end{center}
	\caption{Dependence of opacity and optical depth on $R$ for 
		various AGN disk models and parameters 
		(the inset shows these parameters), of which
		the fiducial values are $M=10^{8}\rm{M}_{\odot}$,  
		$\alpha=0.1$, $\dot{M}=0.1 \dot{M}_{\rm{Edd}}$ 
		adopting SG model. The dashed line in the right 
		panel represents the threshold $\tau=1$. }
	\label{Fig:op}
\end{figure*}

\begin{figure*}
	\begin{center}
		\includegraphics[width=0.45\textwidth]{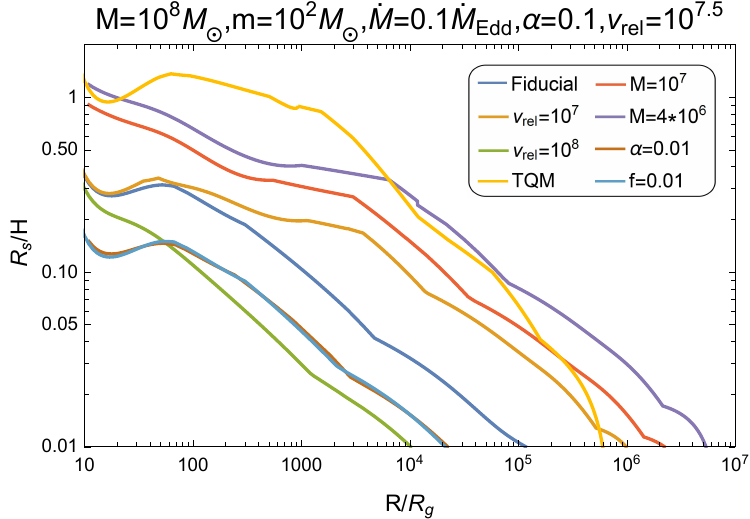}
		\includegraphics[width=0.45\textwidth]{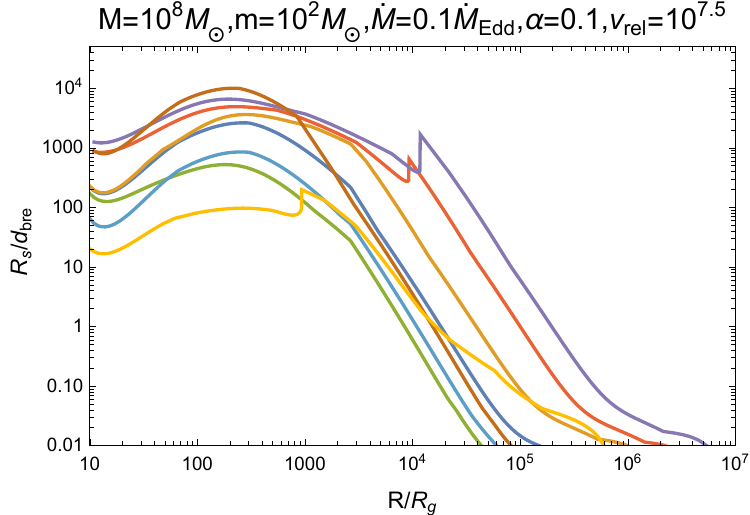}
	\end{center}
	\caption{Comparison of the Strömgren radius with the AGN disk height 
		and the thickness of the jet-head breakout shell for various system 
		parameters. The left and right panel shows 
		$R_{\rm{S}}/H$ and $R_{\rm{S}}/d_{\rm{bre}}$, respectively.}
	\label{Fig:rs}
\end{figure*}

When calculating the structures of AGN disks, 
we have employed the approximate 
expression for the opacity proposed in \cite{Yang19b}; 
the opacity and optical depth properties 
of AGN disks are shown in 
Figure \ref{Fig:op}. $\kappa_{\rm{AGN}}$ 
and $\tau_{\rm{AGN}}$ 
drop sharply at $O(10^3)R_{\rm{g}}$ and 
$>O(10^4)R_{\rm{g}}$ for TQM and SG 
models, by reason of the gas recombination 
at disk temperature 
ranging $10^3-10^4\K$ \citep[e.g.][]{Thompson05}. 
Thus, the AGN disk is optically thin to its 
own radiation at these regions, but it 
would still be opaque to photons with 
energy $>13.6 ~\rm{eV}$ (the ionization energy 
of Hydrogen atom) because of 
the absorption by neutral gas \citep{Gilbaum22}.
As shown in Figure \ref{Fig:rad}, photons 
generated by the BH-driven jet-cocoon 
systems always have characteristic temperatures 
$>10^5\K$, which would thereby ionize 
the neutral gas and cannot directly escape the 
initially \textquotedblleft optically 
thin\textquotedblright AGN disk.

To specifically analyze the propagation of jet 
photons, we estimate the maximum ionization 
distance for the jet-head shock photons with temperature of 
Equation (\ref{Thbre}), via 
the Strömgren radius \citep{Dyson97},
\begin{equation}
R_{\rm{S}}=\left(\frac{3}{4 \pi} 
\frac{S}{n_{\rm{AGN}}^{2} \beta_{2}}\right)^{1 / 3},
\end{equation}  
where $S=L_{\rm{h,bre}}/(k T_{\rm{h,bre}})$ 
is the photon emission rate, 
$n_{\rm{AGN}}=\rho_{\rm{AGN}}/m_{\rm{p}}$, 
and $\beta_{2}=2 \times 
10^{-10}T_{\rm{h,bre}}^{-3 / 4}\rm{~cm}^{3}\rm{~s}^{-1}$. 
As shown in Figure \ref{Fig:rs}, we find that 
$R_{\rm{S}}<H$, which means that 
the environment exhibits high opacity and 
photons cannot escape from the jet-cocoon system, 
maintaining the radiation-dominated jet head 
and cocoon; we also find that 
$R_{\rm{S}}>d_{\rm{bre}}$, thereby the ambient 
gas above the jet head has been ionized at 
its breakout and our assumption of 
$\kappa_{\rm{AGN}}=0.34\cm^2\g^{-1}$ holds except 
for the very outer AGN disk regions, where the 
BH-driven jet is too weak to ionize the medium.

\end{appendix}

\end{document}